\documentclass[lournal]{IEEEtran}

\usepackage{amsmath,graphicx}
\usepackage{amsfonts}  
\usepackage{color}
\usepackage{amssymb}
\usepackage{amsthm}
\usepackage{cite}
\usepackage{array}

\usepackage{dsfont}
\usepackage{algorithm, algorithmic}
\usepackage{bm}
\ifCLASSINFOpdf
\else
\fi
\newcommand{\be}{\begin{equation}}
\newcommand{\ee}{\end{equation}}
\newcommand{\ist}{\hspace*{.3mm}}
\newcommand{\rmv}{\hspace*{-.3mm}}
\newcommand{\nn}{\nonumber}

\allowdisplaybreaks

\begin{document}

\title{A Distributed Particle-PHD Filter with 
Arithmetic-Average PHD Fusion\vspace{2mm}}

\author{Tiancheng~Li and
        Franz~Hlawatsch,~\IEEEmembership{Fellow,~IEEE}
\thanks{T.\ Li is with the School of Automation, Northwestern Polytechnical University, Xi'an 710129, China and also with the BISITE group, Uni\-versity of Salamanca, 37007 Salamanca, Spain; e-mail: t.c.li@usal.es, t.c.li@mail.nwpu. edu.cn.}
\thanks{F.\ Hlawatsch is with the Institute of Telecommunications, TU Wien, Vienna, Austria and also with Brno University of Technology, Brno, Czech Republic; 
e-mail: franz.hlawatsch@tuwien.ac.at.}
\thanks{This work was partially supported by the Marie Sk\l{}odowska-Curie Individual Fellowship (H2020-MSCA-IF-2015) under Grant 709267 and by the Austrian Science Fund (FWF) 
under Grant P27370-N30.}} 
%
%

\markboth{T. Li and F. Hlawatsch, Submitted to IEEE Trans. Signal Process., Dec. 2018}{} 
%



\maketitle

\begin{abstract} 
We propose a particle-based distributed PHD filter for tracking an unknown, time-varying number of targets.
To reduce communication, 
the local PHD filters at neighboring sensors\linebreak 
communicate Gaussian mixture (GM) parameters.
In contrast to most existing distributed PHD filters, 
our
filter employs an ``arithmetic average'' fusion. 
For particles--GM conversion,
we use a method that 
avoids particle clustering and enables \nolinebreak 
a \nolinebreak 
significance-based \nolinebreak 
pruning of the GM components. 
For GM--particles conversion,
we develop an importance sampling based
method 
that enables 
a parallelization of filtering and dissemination/fusion operations. 
The proposed 
distributed particle-PHD filter is able to 
integrate GM-based local PHD filters.
Simulations 
demonstrate the excellent performance and small communication and computation requirements of our filter.
\end{abstract}


%
\IEEEpeerreviewmaketitle

\section{Introduction}

%
%
%
%


\IEEEPARstart{T}{he} probability hypothesis density (PHD) filter is a popular method for tracking an unknown, time-varying number of targets in the presence of
clutter and missed detections \cite{Mahler03,Vo05,Vo06}. In decentralized sensor networks, a distributed extension of the PHD filter can be employed
where each sensor 
runs a local PHD filter and exchanges relevant information with 
neighboring sensors.
For the local PHD filters, a Gaussian mixture (GM) implementation \cite{Battistelli13, Battistelli15, Gunay16, Yi17,Yu16,Li17merging,Li17PC} or a particle-based implementation \cite{Uney10,Uney13,Gostar17,Li18diffusion}
is typically used. 
For distributed data fusion, 
most existing distributed PHD filters perform a ``geometric average'' (GA) fusion of the local posterior PHDs \cite{Uney10,Uney13,Battistelli13, Battistelli15,Gunay16, Yi17}; this type of fusion is also known as 
(generalized) covariance intersection 
\cite{Uhlmann95, Julier01, Bailey12, Mahler00,Clark10, Mahler12}. 
However, GA fusion of PHDs has been observed to suffer from certain deficiencies:
it performs poorly in the case of closely spaced targets \cite{Li17merging,Li17PC}; 
it incurs a delay in detecting new targets \cite{Gunay16,Li17PC}; 
it is sensitive to missing measurements \cite{Yu16, Yi17}; 
and it does not lead to consistent fusion of cardinality distributions 
and thus tends to
underestimate the number of targets \cite{Uney18}. 





In this paper, we propose a distributed PHD filter method that performs an ``arithmetic average'' (AA) fusion of the local\linebreak 
posterior PHDs. 
AA fusion of PHDs 
first appeared indirectly in the context of centralized multisensor PHD filtering,
as an implicit consequence of AA fusion of the generalized \nolinebreak 
likelihood functions of multiple sensors \cite{Streit08}.
It was used explicitly and in the context of distributed PHD filtering in  
\cite{Yu16,Li17PC} (based on a GM implementation of the local PHD filters) and in \cite{Gostar17,Li18diffusion} 
(based on a particle implementation of the local PHD filters and a straightforward particle-based dissemination/fusion scheme). 
AA fusion of PHDs was 
demonstrated 
in \cite{Li17PC,Gostar17,Li18diffusion} to outperform GA fusion of PHDs in the sense of better filtering accuracy, higher reliability in scenarios with strong clutter and/or frequent missed detections, 
and lower computational complexity. 

The proposed distributed PHD filter employs a particle \nolinebreak 
implementation \nolinebreak 
of the local PHD filters for the sake of maximum suitability for nonlinear and/or 
non-Gaussian system models. Straightforward fusion of particle representations of the local and fused PHDs 
imposes high communication requirements \cite{Gostar17,Li18diffusion}. 
By contrast, our filter has low communication \nolinebreak 
requirements
because 
GM parameters are communicated.
This also allows our particle-based local PHD filters to be easily combined with GM-based local PHD filters within a heterogeneous 
network architecture. 

For converting particle representations into GM representations, we propose a data-driven method that avoids
a clustering of the particles. This method generates from the particle representation one Gaussian component for each measurement that has a significant impact on the particle weights.
The overall approach is inspired by a scheme for estimate extraction proposed in \cite{Zhao10,Ristic10,Schikora12}.
For converting
the GMs 
produced by AA fusion into particle representations,
we propose a 
method that is based on importance sampling (IS)
\cite[Ch. 3.3]{Robert05}. This method does not require 
sampling from the fused GM, thereby enabling a parallelization of filtering and dissemination/fusion
operations.
This allows more dissemination/fusion
iterations to be performed 
compared to protocols where the filtering and dissemination/fusion
operations
must be performed serially. 
Overall, the main contribution of this paper is to devise an AA fusion-based distributed particle-PHD filter that has low communication requirements
and allows for a parallelization of filtering and dissemination/fusion operations. 

The paper is organized as follows.
The system model \nolinebreak 
is \nolinebreak 
described in Section \ref{sec:syst}.
Section \ref{sec:phd} discusses the basic \nolinebreak 
operation of the particle-based local PHD filters and presents a measurement-based weight decomposition.
Section \ref{sec:motivation-outline} provides a motivation and outline of the proposed distributed PHD filter.
Section \ref{sec:P2GM} describes a method for converting particle representations into GM representations.
Section \ref{sec:GM-cons} discusses two schemes for GM dissemination and fusion.
An IS method for converting the fused GM into a particle representation is proposed in Section \ref{sec:IS}.
Section \ref{sec:rest} presents two further stages of the proposed distributed PHD filter.
Section \ref{sec:summ_parallel_comm} provides a summary of the overall method,
discusses the parallelization of filtering and fusion, and
analyzes the 
communication cost. Simulation results are reported in Section \ref{sec:simulation}. 

\section{System Model}
\label{sec:syst}

We consider $N_k$ targets with random states $\mathbf{x}_k^{(\nu)}\!\!\in \! \mathbb{R}^d\rmv$, $\nu = 1,2,\ldots,N_k$ at discrete time $k$. 
The number of targets, $N_k$, is unknown, time-varying, and considered random. Accordingly,
the collection of target states 
is modeled by a {random finite set} (RFS) $X_k = \big\{\mathbf{x}_k^{(1)},\mathbf{x}_k^{(2)},\ldots,\mathbf{x}_k^{(N_k)}\big\}$ 
with random cardinality $N_k \!=\rmv |X_k |$ \cite{Mahler07book}. 
The \emph{cardinality distribution} $\rho(n) \triangleq \mathrm{Pr}[N_k \!=\! n]$ 
is the probability mass function of $N_k$.
A target with state $\mathbf{x}_{k-1}$ at time $k \!-\! 1$ continues to exist at time $k$ with 
probability $p_k^{\text{S}} (\mathbf{x}_{k-1})$ (``survival probability'')
or disappears with probability $1 \!-\rmv p_k^{\text{S}} (\mathbf{x}_{k-1})$. In the former case, its new state $\mathbf{x}_k \!\in\! X_k$ is distributed according to a transition 
probability density function (pdf) $f_k(\mathbf{x}_k|\mathbf{x}_{k-1})$. 
There may also be newborn targets, whose states 
are modeled by a Poisson RFS with intensity function $\gamma_k(\mathbf{x}_k)$ \cite{Singh09}.

There are $S$ sensors indexed by $s \rmv\in\rmv \{1,2,\ldots,S\}$.
At time $k$, each sensor $s$ observes an
RFS of measurements $Z_{s,k} \rmv=$\linebreak 
$\big\{{\mathbf{z}_{s,k}^{(1)},\ldots,\mathbf{z}_{s,k}^{(M_{s,k})}}\big\}$, 
where $M_{s,k}$ is the number of measurements observed by sensor $s$ at time $k$. 
We denote by ${\cal S}_s \rmv\subseteq$\linebreak 
$\{1,2,\ldots,S\} \rmv\setminus\rmv \{s\}$ the set of sensors that are connected to sensor $s$ by a communication link,
and we refer to these sensors as the {neighbors} of sensor $s$. We assume that the sensor network is connected, i.e., each sensor can be reached from each 
other sensor by one or multiple communication hops.
A target with state $\mathbf{x}_k$ is ``detected'' by sensor $s$ with probability $p_{s,k}^\text{D}(\mathbf{x}_k)$ (``detection probability'') or ``missed'' by sensor $s$ 
with probability $1 \!-\rmv p_{s,k}^\text{D}(\mathbf{x}_k)$. In the former case, the target generates a measurement $\mathbf{z}_k \!\in\! Z_k$, which is distributed according to the pdf
$g_{s,k}(\mathbf{z}_k \big|\mathbf{x}_k)$. 
There may also be clutter measurements, which are modeled by a Poisson RFS with intensity function (PHD) $\kappa_{s,k}(\mathbf{z}_k)$.
The multitarget state evolution and measurement processes are assumed to satisfy
the independence assumptions
discussed in \cite{Mahler03}.

\section{Local Particle-PHD Filters}
\label{sec:phd}

Each sensor runs a local PHD filter 
that uses the local measurement set $Z_{s,k}$ and communicates 
with its neighbors $r \!\in\! {\cal S}_s$ to exchange relevant information. 
%
%
Let us, at first, consider
the local PHD filter without any cooperation.

\vspace{-1mm}

\subsection{Propagation of the Local Posterior PHD} 
\label{sec:prop}

The local PHD filter propagates the local posterior PHD over time $k$. 
Let $Z_{s,1:k} = (Z_{s,1}, \ldots, Z_{s,k})$ comprise
the local measurements $Z_{s,k'}$ observed by sensor $s$ up to time $k$.
Furthermore, for a region $\mathcal{R} \!\subseteq\! \mathbb{R}^{d}$, let $N_k^\mathcal{R} \triangleq \ist | X_k \!\cap\rmv \mathcal{R} |$ denote the number 
of those targets whose states are in $\mathcal{R}$. Then, the local posterior PHD at sensor $s$, $D_{s,k}(\mathbf{x}|Z_{s,1:k})$, is 
defined as the 
function of $\mathbf{x} \!\in \! \mathbb{R}^d$ 
whose integral over a region $\mathcal{R} \!\subseteq\! \mathbb{R}^{d}$ equals the posterior expectation of $N_k^\mathcal{R}$, i.e.\ \cite{Mahler07book} 
\be
\int_{\mathcal{R}} \!\rmv D_{s,k}(\mathbf{x}|Z_{s,1:k}) \ist \mathrm{d}\mathbf{x} = \mathrm{E}\big[N_k^\mathcal{R}  \big| Z_{s,1:k}\big] \ist.
\label{eq:Cardinality_R}
\ee
In particular, for $\mathcal{R} \!=\! \mathbb{R}^{d}$, we have $N_k^{\mathbb{R}^{d}} \!= | X_k \cap \mathbb{R}^{d} | = | X_k | = N_k$, and thus \eqref{eq:Cardinality_R} becomes
\begin{align}
\int_{\mathbb{R}^d} \!\rmv D_{s,k}(\mathbf{x}|Z_{s,1:k}) \ist \mathrm{d}\mathbf{x} 
= \mathrm{E} [N_k|Z_{s,1:k}] = \sum_{n=0}^\infty \! n \, \rho(n|Z_{s,1:k})\ist, \nn\\[-2.5mm]
\label{eq:Cardinality}\\[-8mm]
\nn
\end{align}
where $\rho(n|Z_{s,1:k}) = \mathrm{Pr}[N_k \!=\! n|Z_{s,1:k}]$.
The posterior expectation of $N_k$, $\mathrm{E} [N_k|Z_{s,1:k}]$, is equal to the minimum mean square error (MMSE) estimate of $N_k$ from 
$Z_{s,1:k}$ \cite{Kay93}, denoted $\hat{N}_{s,k}^{\text{MMSE}}\rmv$. Thus, Eq. \eqref{eq:Cardinality} implies
\be 
\label{eq:Cardinality_MSSE}
\hat{N}_{s,k}^{\text{MMSE}} = \mathrm{E} [N_k|Z_{s,1:k}] =\rmv \int_{\mathbb{R}^d} \!\rmv D_{s,k}(\mathbf{x}|Z_{s,1:k}) \ist \mathrm{d}\mathbf{x}\ist. 
\ee
This is also known as the expected a posteriori (EAP) estimate of $N_k$ \cite{Mahler03,Mahler07book}. 




The local PHD filter 
performs a 
time-recursive calculation of an approximation $\hat{D}_{s,k}(\mathbf{x}|Z_{s,1:k})$ of the local posterior PHD $D_{s,k}(\mathbf{x}|Z_{s,1:k})$.
In a \textit{prediction step},
it converts the preceding approximate local posterior PHD $\hat{D}_{s,k-1}(\mathbf{x}|Z_{s,1:k-1})$ into a ``predicted'' PHD, 
denoted $D_{s,k|k-1}(\mathbf{x}|Z_{s,1:k-1})$, 
via an expression involving $f_k(\mathbf{x}_k|\mathbf{x}_{k-1})$, $p_{k}^{\text{S}}(\mathbf{x})$, and $\gamma_{k}(\mathbf{x})$ \cite{Mahler03}. 
In a subsequent 
\textit{update step}, it converts 
$D_{s,k|k-1}(\mathbf{x}|Z_{s,1:k-1})$ into the 
approximate local posterior PHD $\hat{D}_{s,k}(\mathbf{x}|Z_{s,1:k})$ via an\linebreak 
expression involving $g_{s,k}(\mathbf{z} \big|\mathbf{x})$, 
$p_{s,k}^\text{D}(\mathbf{x})$, and $\kappa_{s,k}(\mathbf{z})$ \cite{Mahler03,Vo05}. 


\vspace{-1mm}

\subsection{Particle-Based Implementation} 
\label{sec:part}

We 
use the particle-based implementation of the prediction and update steps proposed in \cite{Vo05}. 
The approximate local posterior PHD $\hat{D}_{s,k}(\mathbf{x}|Z_{s,1:k})$ is represented by the weighted particle set 
$\xi_{s,k} \rmv\triangleq\rmv \big\{ \big( \mathbf{x}_{s,k}^{(j)} \ist, w_{s,k}^{(j)}\big) \big\}_{j=1}^{J_{s,k}}$, which consists of $J_{s,k}$ particles 
$\mathbf{x}_{s,k}^{(j)} \!\in\!\mathbb{R}^d$ and 
weights $w_{s,k}^{(j)} \!\ge\!0$, $j=1, 2, \ldots, J_{s,k}$. 
The sum of the 
\vspace{-1mm}
weights,
\begin{equation}
\label{eq:W_k:z_def}
W_{s,k} 
\triangleq \sum_{j=1}^{J_{s,k}} w_{s,k}^{(j)} \ist,
\end{equation}
approximates 
$\int_{\mathbb{R}^d} \rmv\hat{D}_{s,k}(\mathbf{x}|Z_{s,1:k}) \ist \mathrm{d}\mathbf{x}$ and, hence,
$\int_{\mathbb{R}^d} \rmv D_{s,k}(\mathbf{x}|Z_{s,1:k}) \ist \mathrm{d}\mathbf{x}$. Thus, it further follows with \eqref{eq:Cardinality_MSSE} \nolinebreak 
that 
\begin{equation}
W_{s,k} \approx 
\hat{N}_{s,k}^{\text{MMSE}} .
\label{eq:W_k:j}
\end{equation}

Propagating 
the approximate local posterior PHD 
(i.e., $\hat{D}_{s,k-1}(\mathbf{x}|Z_{s,1:k-1}) \!\to\! \hat{D}_{s,k}(\mathbf{x}|Z_{s,1:k})$) is now approximated by propagating the weighted particle set, i.e., 
$ \xi_{s,k-1} \!\to\! \xi_{s,k}$. 
The time-recursive calculation of 
$\xi_{s,k}$ is done as follows \cite{Vo05}. 
For each previous particle $\mathbf{x}_{s,k-1}^{(j)}$, 
$j \rmv\in\rmv \{1,\ldots,J_{s,k-1}\}$, a current particle $\mathbf{x}_{s,k}^{(j)}$ is drawn from a proposal pdf 
$q_{s,k}\big(\mathbf{x} ; \mathbf{x}_{s,k-1}^{(j)},Z_{s,k}\big)$. 
In addition, $L_{s,k} \rmv\triangleq\rmv J_{s,k} \rmv-\rmv J_{s,k-1}$ ``newborn'' particles $\mathbf{x}_{s,k}^{(j)}$, $j \rmv=\rmv J_{s,k-1}+1,\ldots,J_{s,k}$
are drawn from a proposal pdf 
$p_{s,k}(\mathbf{x} ; Z_{s,k})$. Then, for each particle $\mathbf{x}_{s,k}^{(j)}$, 
$j \rmv\in \{1,\ldots,J_{s,k}\}$, a ``predicted'' weight $w_{s,k|k-1}^{(j)}$ is calculated as
\vspace{-1.5mm}
\begin{align} 
w_{s,k|k-1}^{(j)} \!=\rmv \begin{cases}
\displaystyle \frac{ f_{k}\big(\mathbf{x}_{s,k}^{(j)} \big| \mathbf{x}_{s,k-1}^{(j)}\big) \ist w_{s,k-1}^{(j)}}{ q_{s,k}\big(\mathbf{x}_{s,k}^{(j)} ; \mathbf{x}_{s,k-1}^{(j)},Z_{s,k}\big) } \ist, 
   & \!\!\! j \rmv=\rmv 1,\ldots,J_{s,k-1}, \\[5.5mm]
\displaystyle \frac{\gamma_k\big(\mathbf{x}_{s,k}^{(j)}\big)}{L_{s,k} \, p_{s,k}\big(\mathbf{x}_{s,k}^{(j)} ; Z_{s,k}\big)} \ist, 
   & \!\!\! j \rmv=\rmv J_{s,k-1} \!+\!\rmv 1,\ldots,J_{s,k}\ist.
\end{cases} \nn\\[-3mm]
\label{eq:w_k|k-1_j}\\[-7.5mm]
\nn
\end{align}
Note that $J_{s,k} \rmv=\rmv J_{s,k-1} + L_{s,k}$. A simple choice of the first proposal pdf 
is $q_{s,k}\big(\mathbf{x} ; \mathbf{x}_{s,k-1}^{(j)},Z_{s,k}\big)=f_{k}\big(\mathbf{x} \big| \mathbf{x}_{s,k-1}^{(j)}\big)$, in which case 
$w_{s,k|k-1}^{(j)} \!=\rmv w_{s,k-1}^{(j)}$ for $j \rmv=\rmv 1,\ldots,J_{s,k-1}$.

For the calculation of the current weights $w_{s,k}^{(j)}$, $j \rmv=\rmv 1,\ldots,$\linebreak 
$J_{s,k}$, we 
formally introduce a ``pseudo-measurement'' $\mathbf{z}_0$ 
representing the case of a missed detection at sensor $s$, and, accordingly, we consider an \emph{extended measurement set} 
$Z_{s,k}^{0} \!\triangleq$\linebreak 
$\{\mathbf{z}_0\} \hspace{.15mm}\cup\hspace{.1mm} Z_{s,k} \!=\! \big\{\mathbf{z}_0, {\mathbf{z}_{s,k}^{(1)},\ldots,\mathbf{z}_{s,k}^{(M_{s,k})}}\big\}$. 
Then, the weight \nolinebreak 
expression in \cite[Eq.\ (22)]{Vo05} can be formulated as the 
\vspace{.3mm}
sum \cite{Zhao10,Ristic10,Schikora12}
\begin{equation}
\label{eq:w_k(j)}
w_{s,k}^{(j)} =\! \sum_{\mathbf{z} \in Z_{s,k}^{0}} \!\rmv\omega_{s,k}^{(j)}(\mathbf{z}) \ist , \quad j \rmv=\rmv 1,\ldots,J_{s,k} \ist,
\vspace{-2.5mm}
\end{equation}
where 
\be
\hspace{1mm}\omega_{s,k}^{(j)}(\mathbf{z}) = \begin{cases}
\big(1 \!-\rmv p_{s,k}^\text{D}\big(\mathbf{x}_{s,k}^{(j)}\big)\big) \ist w_{s,k|k-1}^{(j)} \ist, 
   &\!
   \mathbf{z} \rmv=\rmv \mathbf{z}_0 \\[2.5mm]
\displaystyle \frac{p_{s,k}^\text{D}\big(\mathbf{x}_{s,k}^{(j)}\big) \ist g_{s,k}\big(\mathbf{z} \big| \mathbf{x}_{s,k}^{(j)}\big) \ist w_{s,k|k-1}^{(j)}}{\kappa_{s,k}(\mathbf{z}) 
   + G_{s,k}(\mathbf{z})} \ist, 
   &\!
   \mathbf{z} \rmv\in\! Z_{s,k} \ist,
\end{cases}
\hspace{-5mm}\label{eq:w_decomposition}
\ee
with
$G_{s,k}(\mathbf{z}) \triangleq \sum_{j=1}^{J_{s,k}} p_{s,k}^\text{D}\big(\mathbf{x}_{s,k}^{(j)}\big) \ist g_{s,k}\big(\mathbf{z} \big| \mathbf{x}_{s,k}^{(j)}\big) \ist w_{s,k|k-1}^{(j)}$. 
Expression \eqref{eq:w_k(j)} provides an expansion of 
$w_{s,k}^{(j)}$ into $|Z_{s,k}^{0}| \rmv= M_{s,k}+ 1$ components $\omega_{s,k}^{(j)}(\mathbf{z})$, 
each of which corresponds to one of the measurements $\mathbf{z} \!\in\! Z_{s,k}^{0}$
. We also introduce
\vspace{-.5mm}
\begin{equation}
\Omega_{s,k}(\mathbf{z}) \ist\triangleq\ist \sum_{j=1}^{J_{s,k}} \omega_{s,k}^{(j)}(\mathbf{z}) \ist, \quad \mathbf{z} \rmv\in\! Z_{s,k}^0 \ist.  
\label{eq:W_k(z)}
\vspace{-.5mm}
\end{equation}
For $\mathbf{z} \rmv\in\! Z_{s,k}$, 
$\Omega_{s,k}(\mathbf{z}) \rmv=\rmv 
G_{s,k}(\mathbf{z})/\big( \kappa_{s,k}(\mathbf{z}) + G_{s,k}(\mathbf{z}) \big) \rmv\in\rmv [0,1]$, 
which provides an estimate of the probability that measurement $\mathbf{z}$ originates from a target.
For $\mathbf{z} \rmv=\rmv \mathbf{z}_0$, $\Omega_{s,k}(\mathbf{z}_0) \rmv=  \sum_{j=1}^{J_{s,k}} \big(1 \!-\rmv p_{s,k}^\text{D}\big(\mathbf{x}_{s,k}^{(j)}\big)\big) \ist w_{s,k|k-1}^{(j)}$ 
provides an estimate of the  number of missed detections. Note that
\begin{equation}
\label{eq:W_k_Omega}
\sum_{\mathbf{z} \in Z_{s,k}^{0}} \!\! \Omega_{s,k}(\mathbf{z}) = \sum_{j=1}^{J_{s,k}} \sum_{\mathbf{z} \in Z_{s,k}^{0}} \!\! \omega_{s,k}^{(j)}(\mathbf{z})
= \sum_{j=1}^{J_{s,k}} w_{s,k}^{(j)} = W_{s,k} \ist,
\end{equation}
where \eqref{eq:w_k(j)} and \eqref{eq:W_k:z_def} were used.



\section{Motivation and Outline of the Proposed\vspace{-.3mm}
PHD Fusion Scheme} 
\label{sec:motivation-outline}


The proposed distributed PHD filter uses information fused among the sensors
to ``re-weight'' the particles of the local PHD filters such that the resulting new PHD approximates the AA of the local PHDs. 
Forming the AA 
can be motivated as follows. Suppose 
sensor $s$ wishes to estimate the number of targets in a region $\mathcal{R} \!\subseteq\! \mathbb{R}^{d}$, $N_k^\mathcal{R} \rmv=\rmv | X_k \cap \mathcal{R} |$, 
via the estimator (cf.\ \eqref{eq:Cardinality_R}) $\hat{N}_{s,k}^{\mathcal{R},\text{loc}} \rmv=\rmv \int_{\mathcal{R}} \rmv \hat{D}_{s,k}(\mathbf{x}|Z_{s,1:k}) \ist \mathrm{d}\mathbf{x}$. 
Since 
$Z_{s,1:k}$ is affected by clutter and missed detections, $\hat{N}_{s,k}^{\mathcal{R},\text{loc}}$ may be quite different from $N_k^\mathcal{R}$. 
For example, if one target is in $\mathcal{R}$, i.e., $N_k^\mathcal{R} \!=\! 1$, sensor $s$ may fail to detect that target, resulting in 
$\hat{N}_{s,k}^{\mathcal{R},\text{loc}} \!\rmv\approx\rmv 0$; or if no target is in $\mathcal{R}$, i.e., $N_k^\mathcal{R} \!=\rmv 0$, a false alarm (clutter) at sensor $s$ 
may lead to $\hat{N}_{s,k}^{\mathcal{R},\text{loc}} \!\rmv\approx\! 1$. On the other hand, because the clutter and the missed detections at different sensors $s \!\in\! \cal{S}$
are not identical---in fact, they are assumed 
independent across the sensors---one can expect that the AA of the $\hat{N}_{s,k}^{\mathcal{R},\text{loc}}\rmv$, 
$\hat{N}_{1:S,k}^\mathcal{R} \rmv\triangleq \sum_{s=1}^S \rmv \hat{N}_{s,k}^{\mathcal{R},\text{loc}} / S$, is a more robust
estimate of $N_k^\mathcal{R}$. This AA can be expressed 
\vspace{-.5mm}
as
\[
\hat{N}_{1:S,k}^\mathcal{R} = \frac{1}{S} \sum_{s=1}^S \int_{\mathcal{R}} \!\! \hat{D}_{s,k}(\mathbf{x}|Z_{s,1:k}) \ist \mathrm{d}\mathbf{x}
  =\rmv \int_{\mathcal{R}} \!\! \hat{D}_{k}(\mathbf{x}|Z_{1:S,1:k}) \ist \mathrm{d}\mathbf{x} \ist,
\vspace{.5mm}
\]
with the AA of the local PHDs
\be
\hat{D}_{k}(\mathbf{x}|Z_{1:S,1:k}) \triangleq \frac{1}{S} \sum_{s=1}^S \hat{D}_{s,k}(\mathbf{x}|Z_{s,1:k}) \ist.
\label{eq:av-local-PHDs}
\vspace{-.5mm}
\ee
Thus, 
$\hat{N}_{1:S,k}^\mathcal{R}$ is obtained by integrating the AA of the local PHDs over $\mathcal{R}$.
This motivates a fusion of the local PHDs $\hat{D}_{s,k}(\mathbf{x}|Z_{s,1:k})$---thereby combining all the local measurements $Z_{s,1:k}$, $s=1,\ldots,S$---by calculating the 
AA of the $\hat{D}_{s,k}(\mathbf{x}|Z_{s,1:k})$: we can expect that this
compensates the effects of clutter and missed detections to some extent. 
In addition, the AA fusion of the local PHDs can be motivated theoretically by the fact that 
the fused PHD minimizes the sum of the Cauchy-Schwarz divergences relative to the local PHDs \cite{Hoang15,Gostar17}. 



To reduce the amount of intersensor communication,
the information exchanged between neighboring sensors in our approach consists of GM parameters rather than particles and weights. 
This 
necessitates conversions between particle 
and GM representations. The proposed AA-based fusion scheme thus consists of the following 
\vspace{.5mm}
steps:

\begin{enumerate}

\item Each sensor $s$ converts its 
weighted particle set $\xi_{s,k} = \big\{ \big( \mathbf{x}_{s,k}^{(j)} \ist, w_{s,k}^{(j)}\big) \big\}_{j=1}^{J_{s,k}}$ into a GM (see Section \ref{sec:P2GM})
and broadcasts the GM parameters to the neighboring sensors $r \!\in\! {\cal S}_s$.

\vspace{.7mm}

\item Each sensor $s$ broadcasts its local cardinality estimate $W_{s,k}$ (see \eqref{eq:W_k:z_def}, \eqref{eq:W_k:j})
to the neighboring sensors $r \!\in\! {\cal S}_s$.

\vspace{.7mm}

\item The 
GM parameters of each sensor $s$ are fused with those received from the other sensors via
a distributed dissemination/fusion scheme;
see Section \ref{sec:GM-cons}.

\vspace{.7mm}

\item The local cardinality estimate $W_{s,k}$ of each sensor $s$ is fused with those received from the other sensors via
a distributed dissemination/fusion scheme; 
see Section \ref{sec:CC-AA} \cite{Li19CC}.

\vspace{.7mm}

\item At each sensor $s$, the local particle weights $w_{s,k}^{(j)}$ are modified based on the fused GM parameters and the fused cardinality estimate; see Sections
\ref{sec:IS} and \ref{sec:CC-AA}.


\end{enumerate}




\section{Particles--GM Conversion} 
\label{sec:P2GM}

In Step 1, 
the local weighted particle set $\xi_{s,k} = \big\{ \big( \mathbf{x}_{s,k}^{(j)} \ist,$\linebreak 
$w_{s,k}^{(j)}\big) \big\}_{j=1}^{J_{s,k}}$ is converted into a GM representation.
Our conversion method differs from previous
methods
\cite{Coates04, Sheng05, Gu07, Hlinka13, Uney13, Li17flooding, Li18Nehorai}\linebreak 
in that it is based on the weight expansion in \eqref{eq:w_k(j)}, i.e., 
$w_{s,k}^{(j)} =$\linebreak 
$\sum_{\mathbf{z} \in Z_{s,k}^{0}} \rmv\omega_{s,k}^{(j)}(\mathbf{z})$.
In our method, each of the $|Z_{s,k}^{0}| \rmv=\rmv M_{s,k} \rmv+\rmv 1$\linebreak 
extended measurements 
$\mathbf{z} \rmv\in\rmv Z_{s,k}^{0} \!=\rmv \big\{\mathbf{z}_0, \mathbf{z}_{s,k}^{(1)},\ldots,\mathbf{z}_{s,k}^{(M_{s,k})}\big\}$
potentially corresponds to one Gaussian component (GC) $\mathcal{N}\big(\mathbf{x};\bm{\mu}_{s,k}(\mathbf{z}), \bm{\Sigma}_{s,k}(\mathbf{z})\big)$.
Here, $\mathcal{N}(\mathbf{x};\bm{\mu},\bm{\Sigma})$ denotes a Gaussian pdf with mean vector $\bm{\mu}$ and covariance matrix $\bm{\Sigma}$. The GC
$\mathcal{N}\big(\mathbf{x};\bm{\mu}_{s,k}(\mathbf{z}), \bm{\Sigma}_{s,k}(\mathbf{z})\big)$
is meant to represent the weighted particle set $\big\{ \big(\mathbf{x}_{s,k}^{(j)}\ist , \omega_{s,k}^{(j)}(\mathbf{z}) \big)\big\}_{j=1}^{J_{s,k}}$.
The mean vector $\bm{\mu}_{s,k}(\mathbf{z})$ and covariance matrix 
$\bm{\Sigma}_{s,k}(\mathbf{z})$ are derived from the respective weight components $\omega_{s,k}^{(j)}(\mathbf{z})$
and the particles
$\mathbf{x}_{s,k}^{(j)}$ 
\vspace{-.4mm}
as
\begin{align} 
\bm{\mu}_{s,k}(\mathbf{z}) &= \sum_{j=1}^{J_{s,k}} \bar{\omega}_{s,k}^{(j)}(\mathbf{z}) \, \mathbf{x}_{s,k}^{(j)} ,
\label{eq:GC_mean} \\[.5mm]
\bm{\Sigma}_{s,k}(\mathbf{z}) &= \sum_{j=1}^{J_{s,k}} \bar{\omega}_{s,k}^{(j)}(\mathbf{z}) \ist \big( \mathbf{x}_{s,k}^{(j)} \rmv-\rmv \bm{\mu}_{s,k}(\mathbf{z}) \big) \ist
  \big( \mathbf{x}_{s,k}^{(j)} \rmv-\rmv \bm{\mu}_{s,k}(\mathbf{z}) \big)^{\rmv\text{T}} , \nn\\[-4mm]
\label{eq:GC_P} \\[-7mm]
\nn
\end{align} 
where $\bar{\omega}_{s,k}^{(j)}(\mathbf{z}) = \omega_{s,k}^{(j)}(\mathbf{z}) / \sum_{j'=1}^{J_{s,k}}\omega_{s,k}^{(j')}(\mathbf{z})
= \omega_{s,k}^{(j)}(\mathbf{z}) / \Omega_{s,k}(\mathbf{z})$ with $\omega_{s,k}^{(j)}(\mathbf{z})$ given by \eqref{eq:w_decomposition}.
In the overall GM-based PHD (briefly referred to as GM-PHD), the GC $\mathcal{N}\big(\mathbf{x};\bm{\mu}_{s,k}(\mathbf{z}),$\linebreak 
$\bm{\Sigma}_{s,k}(\mathbf{z})\big)$ is multiplied by the weight $\Omega_{s,k}(\mathbf{z}) = \sum_{j=1}^{J_{s,k}} \omega_{s,k}^{(j)}(\mathbf{z})$ (see \eqref{eq:W_k(z)}).
Thus, there is one weighted GC $\Omega_{s,k}(\mathbf{z}) \, \mathcal{N}\big(\mathbf{x};$\linebreak 
$\bm{\mu}_{s,k}(\mathbf{z}),\bm{\Sigma}_{s,k}(\mathbf{z})\big)$ for each measurement $\mathbf{z} \rmv\in\rmv Z_{s,k}^{0}$. 

The overall GM-PHD is meant to represent the local weighted particle set 
$\xi_{s,k}  = \big\{ \big( \mathbf{x}_{s,k}^{(j)} \ist, w_{s,k}^{(j)}\big) \big\}_{j=1}^{J_{s,k}}$. Because 
$w_{s,k}^{(j)}$\linebreak 
$= \sum_{\mathbf{z} \in Z_{s,k}^{0}} \rmv\omega_{s,k}^{(j)}(\mathbf{z})$, the overall GM-PHD is ideally taken to be the sum of all the weighted GCs, i.e.,
\begin{equation} 
\label{eq:GM-PHD_full}
D_{s,k}^\text{GM,full}(\mathbf{x}) \ist\triangleq\! \sum_{\mathbf{z} \in Z_{s,k}^{0}} \!\! \Omega_{s,k}(\mathbf{z})
\, \mathcal{N}\big(\mathbf{x};\bm{\mu}_{s,k}(\mathbf{z}), \bm{\Sigma}_{s,k}(\mathbf{z})\big) \ist.
\vspace{-.5mm}
\end{equation} 
This provides an approximate GM representation of 
$\hat{D}_{s,k}(\mathbf{x}|Z_{s,1:k})$. 
However, to further reduce the communication cost
,
we 
restrict the sum \eqref{eq:GM-PHD_full} to the GCs corresponding to ``significant'' measurements.
(We note that a similar restriction 
was used previously for estimate extraction in \cite{Zhao10,Ristic10,Schikora12}.) The subset of significant measurements, $Z_{s,k}^\text{S} \!\rmv\subseteq\! Z_{s,k}^{0}$, 
is defined as the set of those $\mathbf{z} \!\in\! Z_{s,k}^{0}$ 
for which $\Omega_{s,k}(\mathbf{z})$ in \eqref{eq:W_k(z)} is above a 
threshold $T_\Omega$, where $0 \!<\! T_\Omega \!<\! 1$.
In other words, the GM at sensor $s$ contains a GC for $\mathbf{z} \rmv\in\! Z_{s,k}$ if the estimated probability that the measurement $\mathbf{z}$ originates from a target 
(given by $\Omega_{s,k}(\mathbf{z})$) is above $T_\Omega$, and it contains a GC for $\mathbf{z}_0$ if the estimated number of missed detections (given by 
$\Omega_{s,k}(\mathbf{z}_0)$) is above $T_\Omega$.
Thus, the local GM-PHD is taken to be
\begin{equation} 
\label{eq:GM-PHD}
D_{s,k}^\text{GM}(\mathbf{x}) \ist\triangleq\! \sum_{\mathbf{z} \in Z_{s,k}^\text{S}} \!\! \Omega_{s,k}(\mathbf{z})
\, \mathcal{N}\big(\mathbf{x};\bm{\mu}_{s,k}(\mathbf{z}), \bm{\Sigma}_{s,k}(\mathbf{z})\big) \ist.
\vspace{-.5mm}
\end{equation}
This can be interpreted as the GM-PHD corresponding to the\linebreak 
particle set $ \big\{ \big( \mathbf{x}_{s,k}^{(j)} \ist, \breve{w}_{s,k}^{(j)}\big) \big\}_{j=1}^{J_{s,k}}$
whose weights $\breve{w}_{s,k}^{(j)}$ are defined by summing the $\omega_{s,k}^{(j)}(\mathbf{z})$ only over the significant measurements, i.e.,
$\breve{w}_{s,k}^{(j)} \!=\! \sum_{\mathbf{z} \in Z_{s,k}^\text{S}} \!\omega_{s,k}^{(j)}(\mathbf{z})$.
We note that an alternative
definition of a significant measurement subset $Z_{s,k}^\text{S}$ and, thus,
of $D_{s,k}^\text{GM}(\mathbf{x})$ is to choose 
the $N_\Omega \rmv\triangleq\rmv \mathrm{round}\ist\{ W_{s,k} \}$ GCs with the largest $\Omega_{s,k}(\mathbf{z})$, $\mathbf{z} \!\in\! Z_{s,k}^{0}$.
Here, 
$W_{s,k} \!=\! \sum_{\mathbf{z} \in Z_{s,k}^{0}} \!\! \Omega_{s,k}(\mathbf{z})$ according to \eqref{eq:W_k_Omega}, and we recall from \eqref{eq:W_k:j} that $W_{s,k}$ 
approximates the MMSE estimate $\hat{N}_{s,k}^{\text{MMSE}}\rmv$.

The suppression of GCs in \eqref{eq:GM-PHD} 
is motivated by the notion that ``insignificant'' measurements are likely to be false alarms (clutter).
However, if an insignificant measurement is not a false alarm after all, we can expect that it is not suppressed at most of the other sensors, and thus the erroneous suppression
at sensor $s$ will be compensated by the subsequent AA fusion. This is an advantage of AA fusion over GA fusion.

The GM parameter set underlying the local GM-PHD $D_{s,k}^\text{GM}(\mathbf{x})$ in \eqref{eq:GM-PHD} is
\begin{equation}
    \label{eq:def_g_s_k}
\mathcal{G}_{s,k} \triangleq \big\{ \big( \Omega_{s,k}(\mathbf{z}), \bm{\mu}_{s,k}(\mathbf{z}), \bm{\Sigma}_{s,k}(\mathbf{z}) \big) \big\}_{\mathbf{z} \in Z_{s,k}^\text{S}} .    
\end{equation}
All the further operations of our
distributed PHD filter are based 
on $\mathcal{G}_{s,k}$; the 
GM-PHD $D_{s,k}^\text{GM}(\mathbf{x})$ itself is never
calculated. These further operations comprise a distributed fusion of the local GM parameter sets 
and of the local cardinality estimates, 
the conversion of the fused GM representations into particle representations, a scaling of the particle weights,
and the calculation of state estimates. A detailed presentation of these steps 
will be given in Sections \ref{sec:GM-cons}--\ref{sec:summ_parallel_comm}.


\section{Two GM Dissemination/Fusion Schemes} 
\label{sec:GM-cons}


Once the local GM parameter sets $\mathcal{G}_{s,k}$ are available at the respective sensors $s$, they are disseminated and fused via a distributed 
scheme. The goal of this 
scheme is to obtain, at each sensor $s$, a GM parameter set that approximately corresponds to 
the AA of all the local GM-PHDs, 
\be 
\bar{D}_{k}^\text{GM}(\mathbf{x}) \triangleq \frac{1}{S} \sum _{s=1}^S D_{s,k}^\text{GM}(\mathbf{x}) \ist.
\label{eq:GM-PHD_AA}
\ee
Note that this equals \eqref{eq:av-local-PHDs} except that 
$\hat{D}_{s,k}(\mathbf{x}|Z_{s,1:k})$ is
replaced by 
$D_{s,k}^\text{GM}(\mathbf{x})$. Next,
we 
discuss two alternative schemes for disseminating and fusing the local GM parameter sets.

\vspace{-1mm}

\subsection{GM Flooding} 
\label{sec:P2GM_flood}

In the flooding scheme \cite{Li17flooding}, each sensor $s$ first broadcasts to its neighbors $r \!\in\! {\cal S}_s$ its GM parameter set
$\mathcal{G}_{s,k} \rmv= \big\{ \big( \Omega_{s,k}(\mathbf{z}),$\linebreak 
$\bm{\mu}_{s,k}(\mathbf{z}), \bm{\Sigma}_{s,k}(\mathbf{z}) \big) \big\}_{\mathbf{z} \in Z_{s,k}^\text{S}}$
and receives 
their GM parameter sets 
$\mathcal{G}_{r,k} \rmv=\rmv \big\{ \big( \Omega_{r,k}(\mathbf{z}), \bm{\mu}_{r,k}(\mathbf{z}),\bm{\Sigma}_{r,k}(\mathbf{z}) \big) \big\}_{\mathbf{z} \in Z_{r,k}^\text{S}}\rmv$, 
$r \!\in\! {\cal S}_s$. Each sensor then augments its own GM parameter set $\mathcal{G}_{s,k}$ by the neighbor GM parameter sets $\mathcal{G}_{r,k}$, $r \!\in\! {\cal S}_s$,
resulting in 
$\mathcal{G}_{s,k}^{\text{F}[1]} \rmv=$\linebreak 
$\bigcup_{r\in \{s\} \cup {\cal S}_s} \rmv \mathcal{G}_{r,k}$. 
In the subsequent flooding iteration $i \rmv\in\rmv \{
2,3,$\linebreak 
$\ldots\}$, each sensor $s$ broadcasts to its neighbors the augmented set $\mathcal{G}_{s,k}^{\text{F}[i-1]}$ with the exception of the elements already broadcast 
(the sensor
keeps track of all the elements it already broadcast \cite{Li17flooding})
and receives the new elements of the neighbors' augmented sets $\mathcal{G}_{r,k}^{\text{F}[i-1]}\rmv$.
This results in the new augmented set
\be 
\label{eq:GM_flooding_i}
\mathcal{G}_{s,k}^{\text{F}[i]} =\! \bigcup_{r\in \{s\} \cup {\cal S}_s} \!\!\! \mathcal{G}_{r,k}^{\text{F}[i-1]} \ist. 
\vspace{-1.5mm}
\ee
This recursion is initialized with $\mathcal{G}_{s,k}^{\text{F}[0]} \!=\rmv \mathcal{G}_{s,k}$.

After the final flooding iteration $i \!=\! I$
(the choice of $I$ will be discussed in Section \ref{sec:summ-parallel}), the augmented parameter set at sensor $s$ is equal to
\be 
\label{eq:GM_flooding}
\mathcal{G}_{s,k}^{\text{F}[I]} 
  =\rmv \bigcup_{r\in {\cal S}_s^{[I]}} \!\! \mathcal{G}_{r,k} \ist, 
\vspace{-1.5mm}
\ee
where ${\cal S}_s^{[I]} \!\subseteq\rmv \{1,2,\ldots,S\}$
denotes the set of all those sensors that are at most $I$ hops away from sensor $s$, including sensor $s$ itself. 
At this point, sensor $s$ would be able to calculate the AA of all the 
GM-PHDs 
whose GM parameters are contained in $\mathcal{G}_{s,k}^{[I]}$, 
\vspace{-2mm}
i.e.,
\be \label{eq:D_flooding}
D_{s,k}^{\text{GM}[I]}(\mathbf{x}) = \frac{1}{\big| {\cal S}_s^{[I]} \big|} \rmv\sum _{r\in {\cal S}_s^{[I]} } \!\! D_{r,k}^{\text{GM}}(\mathbf{x}) \ist.
\ee
If $I \rmv\geq\rmv R$, where $R$ is the network diameter \cite{Olfati-Saber07,Li17flooding}, then
$\mathcal{G}_{s,k}^{\text{F}[I]}$ contains the GM parameters of all the sensors,
and thus $D_{s,k}^{\text{GM}[I]}(\mathbf{x})$ equals the total GM-PHD average
$\bar{D}_{k}^\text{GM}(\mathbf{x})$ in \eqref{eq:GM-PHD_AA}. 
(This presupposes that the sensor network is connected, which we assumed in Section \ref{sec:syst}.)
If $I \rmv<\rmv R$, then $D_{s,k}^{\text{GM}[I]}(\mathbf{x})$ provides only an approximation of $\bar{D}_{k}^\text{GM}(\mathbf{x})$.

A 
drawback of the 
flooding scheme is that as the flooding iteration proceeds, the sets $\mathcal{G}_{s,k}^{\text{F}[i]}$ grow
in size since the GM parameters of additional sensors are included. 
Indeed, in iteration $i$, sensor $s$ receives the GM parameters ${\{\mathcal{G}_{r,k}\}}_{r\in \Delta{\cal S}_s^{[i]}}$, where $\Delta{\cal S}_s^{[i]} \!\subseteq\rmv \{1,2,\ldots,S\}$
comprises
all sensors that are exactly $i$ hops away from sensor $s$; note that $\Delta{\cal S}_s^{[i]} = {\cal S}_s^{[i]} \rmv\setminus {\cal S}_s^{[i-1]}\rmv$. 
These GM parameters are added to the previous GM parameter set of\linebreak 
sensor $s$, $\mathcal{G}_{s,k}^{\text{F}[i-1]}\rmv$. 
Thus, Eq.\ \eqref{eq:GM_flooding_i} can be reformulated as
\be 
\label{eq:GM_flooding_i_2}
\mathcal{G}_{s,k}^{\text{F}[i]} =\ist \mathcal{G}_{s,k}^{\text{F}[i-1]} \cup\! 
\bigcup_{r\in \Delta{\cal S}_s^{[i]}} \!\!\! \mathcal{G}_{r,k} \ist. 
\ee
The total number of 
real values that have to be broadcast in iteration $i$ by all the 
sensors in the network is equal to the number of real values needed to specify 
all the elements of the set 
$\bigcup_{s=1}^S \rmv \bigcup_{r\in \Delta{\cal S}_s^{[i]}} \rmv\mathcal{G}_{r,k}$.

\vspace{-1mm}

\subsection{GM Average Consensus} 
\label{sec:P2GM_merg}

To limit the growth of the GM parameter sets and to reduce the communication cost, we may emulate a part of the averaging in \eqref{eq:D_flooding}
in each iteration $i$.
To this end, we consider a formal application of the average consensus algorithm \cite{Xiao04,Olfati-Saber07} to the local GM-PHDs.
According to that algorithm, the iterated GM-PHD at sensor $s$---denoted by $D_{s,k}^{\text{cons}[i]}(\mathbf{x})$---would be updated in iteration $i$ as
\be 
\label{eq:D_averaging} 
D_{s,k}^{\text{cons}[i]}(\mathbf{x}) = \! \sum _{r\in \{s\} \cup {\cal S}_s } \!\!\!  \alpha_{s,r} \ist D_{r,k}^{\text{cons}[i-1]}(\mathbf{x}) \ist,
\ee  
with appropriately chosen weights $\alpha_{s,r}$, where $s,r \!\in\! \{1,2,$\linebreak 
$\ldots,S\}$.
A popular choice is given by the 
Metropolis weights \cite{Xiao04} defined as $\alpha_{s,r} = 1/ (1+\max{(|{\cal S}_r|,|{\cal S}_s|)})$ if $r \!\neq\! s$ and 
$\alpha_{s,s} = 1 \rmv-\rmv \sum_{r\in {\cal S}_s} \rmv \alpha_{s,r}$.
The recursion \eqref{eq:D_averaging} is initialized as $D_{s,k}^{\text{cons}[0]}(\mathbf{x}) \triangleq D_{s,k}^{\text{GM}}(\mathbf{x})$ (see \eqref{eq:GM-PHD}).
Since the network is connected, 
$D_{s,k}^{\text{cons}[i]}(\mathbf{x})$ is
guaranteed to converge for $i \!\to\! \infty$ to the total GM-PHD average $\bar{D}_{k}^\text{GM}(\mathbf{x})$ in \eqref{eq:GM-PHD_AA} \cite{Xiao04}. 
For a finite number $I$ of iterations, $D_{s,k}^{\text{cons}[I]}(\mathbf{x})$ provides only an approximation of $\bar{D}_{k}^\text{GM}(\mathbf{x})$.

A direct implementation of the update \eqref{eq:D_averaging} is impossible because the iterated GM-PHDs $D_{s,k}^{\text{cons}[i]}(\mathbf{x})$ are functions,
rather than numbers. Therefore, we will emulate \eqref{eq:D_averaging} through operations involving the GM parameters of
the iterated local GM-PHDs $D_{s,k}^{\text{cons}[i]}(\mathbf{x})$ and $D_{r,k}^{\text{cons}[i-1]}(\mathbf{x})$, $r \!\in\! \{s\} \cup {\cal S}_s$ involved in \eqref{eq:D_averaging}.
%
First, as in the 
flooding scheme discussed in Section \ref{sec:P2GM_flood}, each sensor $s$ broadcasts to its neighbors $r \!\in\! {\cal S}_s$ its GM parameter set
$\mathcal{G}_{s,k} \rmv=\! \big\{ \big( \Omega_{s,k}(\mathbf{z}), \bm{\mu}_{s,k}(\mathbf{z}),\bm{\Sigma}_{s,k}(\mathbf{z}) \big) \big\}_{\mathbf{z} \in Z_{s,k}^\text{S}}\!\!$  
(see\linebreak 
\eqref{eq:def_g_s_k}) 
and receives 
their GM parameter sets $\mathcal{G}_{r,k}$. 
Then, sensor $s$ scales each GM weight 
$\Omega_{r,k}(\mathbf{z})$ with the corresponding \nolinebreak 
con\-sensus \nolinebreak 
weight $\alpha_{s,r}$, resulting in the scaled weights 
$\Omega_{s,r,k}^{(\alpha)}(\mathbf{z})$\linebreak 
$\triangleq\rmv \alpha_{s,r} \ist \Omega_{r,k}(\mathbf{z})$, for 
$\mathbf{z} \!\in\! Z_{r,k}^\text{S}$, $r \!\in\! \{s\} \cup {\cal S}_s$. Thus, sensor $s$ now disposes of the ``scaled GM parameter sets'' 
\[
\mathcal{G}_{s,r,k}^{(\alpha)} \rmv\triangleq\rmv \big\{ \big( \Omega_{s,r,k}^{(\alpha)}(\mathbf{z}), \bm{\mu}_{r,k}(\mathbf{z}), \bm{\Sigma}_{r,k}(\mathbf{z}) \big) \big\}_{\mathbf{z} \in Z_{r,k}^\text{S}}\rmv,
\vspace{-.4mm}
\]
for all $r \rmv\in\rmv \{s\} \rmv\cup {\cal S}_s$. The GM-PHD generated in analogy to \eqref{eq:GM-PHD} from the union of all these 
GM parameter sets, 
$\mathcal{G}_{s,k}^{\cup} \!\triangleq \bigcup_{r\in \{s\} \cup {\cal S}_s} \rmv \mathcal{G}_{s,r,k}^{(\alpha)}$, would be 
\begin{align} 
D_{s,k}^{\text{GM},\cup}(\mathbf{x}) &\triangleq\! \sum _{r\in \{s\} \cup {\cal S}_s } \sum_{\mathbf{z} \in Z_{r,k}^\text{S}} \!\! \Omega_{s,r,k}^{(\alpha)}(\mathbf{z})
  \, \mathcal{N}\big(\mathbf{x};\bm{\mu}_{r,k}(\mathbf{z}), \bm{\Sigma}_{r,k}(\mathbf{z})\big) \nn\\[.5mm]
&=\! \sum _{r\in \{s\} \cup {\cal S}_s } \!\!\rmv  \alpha_{s,r} \!\rmv\sum_{\mathbf{z} \in Z_{r,k}^\text{S}} \!\! \Omega_{r,k}(\mathbf{z})
  \, \mathcal{N}\big(\mathbf{x};\bm{\mu}_{r,k}(\mathbf{z}), \bm{\Sigma}_{r,k}(\mathbf{z})\big) \nn\\[.5mm]
&=\! \sum _{r\in \{s\} \cup {\cal S}_s } \!\!\!  \alpha_{s,r} \ist D_{r,k}^\text{GM}(\mathbf{x}) \ist,
\label{eq:GM-PHD_cons-fused}
\end{align} 
where \eqref{eq:GM-PHD} was used in the last step. A comparison with \eqref{eq:D_averaging} shows that we have emulated the first GM-PHD average consensus iteration 
($i \!=\!1$) by operating at the level of the GM parameters \cite{Li17PC}. 
Note, however, that $D_{s,k}^{\text{GM},\cup}(\mathbf{x})$ (or any other PHD) is not actually 
computed by the proposed algorithm.

Just as the flooding scheme, this 
scheme suffers from the fact that the fused GM parameter set at sensor $s$,
$\mathcal{G}_{s,k}^{\cup} \!= \bigcup_{r\in \{s\} \cup {\cal S}_s} \rmv \mathcal{G}_{s,r,k}^{(\alpha)}$, is much larger than the 
original GM parameter set $\mathcal{G}_{s,k}$. Therefore, we apply \emph{mixture reduction}
\cite{Salmond90, Reece10, Li17PC} 
to $\mathcal{G}_{s,k}^{\cup}$, resulting in a reduced GM parameter set 
$\mathcal{G}_{s,k}^{[1]} \rmv\triangleq\rmv \big\{ \big( \Omega_{s,k,\ell}^{[1]}, \bm{\mu}_{s,k,\ell}^{[1]},\bm{\Sigma}_{s,k,\ell}^{[1]} \big) \big\}_{\ell \in L_{s,k}^{[1]}}$, where 
$L_{s,k}^{[1]}$ is some reduced index set. The GM-PHD corresponding to $\mathcal{G}_{s,k}^{[1]}$, i.e.,
\be
D_{s,k}^{\text{GM}[1]}(\mathbf{x}) \triangleq\! \sum_{\ell \in L_{s,k}^{[1]}} \!\! \Omega_{s,k,\ell}^{[1]} 
  \,\mathcal{N}\big(\mathbf{x};\bm{\mu}_{s,k,\ell}^{[1]}, \bm{\Sigma}_{s,k,\ell}^{[1]} \big) \ist,
\vspace{-1.8mm}
\label{eq:GM-PHD_cons-fused_1}
\ee
is then only an approximation of $D_{s,k}^{\text{GM},\cup}(\mathbf{x})$.
Mixture reduction usually consists of merging 
GCs that are ``close'' with respect to an appropriate metric, 
and pruning 
GCs with small weights. In our case, the weights are not small because they survived the thresholding performed in Section \ref{sec:P2GM}, and thus we only perform a merging operation.


These union and 
merging operations are repeated in all the further iterations. In iteration $i \rmv\in\rmv \{2,3,\ldots\}$, sensor $s$ broadcasts 
to its neighbors the set $\mathcal{G}_{s,k}^{[i-1]} = \big\{ \big( \Omega_{s,k,\ell}^{[i-1]}, \bm{\mu}_{s,k,\ell}^{[i-1]},$\linebreak 
$\bm{\Sigma}_{s,k,\ell}^{[i-1]} \big) \big\}_{\ell \in L_{s,k}^{[i-1]}}$ and receives their sets $\mathcal{G}_{r,k}^{[i-1]}\rmv$, $r \!\in\! {\cal S}_s$. 
\vspace{-.7mm}
It then scales each GM weight 
$\Omega_{r,k,\ell}^{[i-1]}$, $\ell \!\in\! L_{r,k}^{[i-1]}\rmv$, $r \!\in\! \{s\} \cup {\cal S}_s$ with the corresponding consensus weight $\alpha_{s,r}$. 
This results in the ``scaled GM parameter sets'' 
\[
\mathcal{G}_{s,r,k}^{[i-1](\alpha)} 
\hspace{-.4mm}\triangleq\rmv \big\{ \big( \Omega_{s,r,k,\ell}^{[i-1](\alpha)}\!, \bm{\mu}_{r,k,\ell}^{[i-1]},\bm{\Sigma}_{r,k,\ell}^{[i-1]} \big) \big\}_{\ell \in L_{r,k}^{[i-1]}},
\; r \!\in\! \{s\} \cup\ist {\cal S}_s \ist,
\]
with 
$\Omega_{s,r,k,\ell}^{[i-1](\alpha)} \rmv\triangleq\rmv \alpha_{s,r} \ist \Omega_{r,k,\ell}^{[i-1]}$. Let
$D_{s,k}^{\text{GM}[i-1],\cup}(\mathbf{x})$ denote the GM-PHD corresponding to the union 
of all these GM parameter sets, $\mathcal{G}_{s,k}^{[i-1],\cup} \!\rmv\triangleq\rmv \bigcup_{r\in \{s\} \cup {\cal S}_s} \rmv \mathcal{G}_{s,r,k}^{[i-1](\alpha)}\rmv$, 
\pagebreak 
i.e.,
\[
D_{s,k}^{\text{GM}[i-1],\cup}(\mathbf{x}) \triangleq\!\! \sum _{r\in \{s\} \cup {\cal S}_s } \sum_{\ell \in L_{r,k}^{[i-1]}} \!\!\! \Omega_{s,r,k,\ell}^{[i-1](\alpha)}
  \ist \mathcal{N}\big(\mathbf{x};\bm{\mu}_{r,k,\ell}^{[i-1]},\bm{\Sigma}_{r,k,\ell}^{[i-1]} \big) \ist.
\vspace{-1.7mm}
\]
Using \eqref{eq:GM-PHD_cons-fused_1} with obvious modifications, i.e., 
$D_{s,k}^{\text{GM}[i-1]}(\mathbf{x}) = \sum_{\ell \in L_{s,k}^{[i-1]}} \Omega_{s,k,\ell}^{[i-1]} 
  \,\mathcal{N}\big(\mathbf{x};\bm{\mu}_{s,k,\ell}^{[i-1]}, \bm{\Sigma}_{s,k,\ell}^{[i-1]} \big)$, we 
obtain 
\vspace{-1mm}
(cf.\ \eqref{eq:GM-PHD_cons-fused})
\begin{equation} 
D_{s,k}^{\text{GM}[i-1],\cup}(\mathbf{x}) 
=\! \sum _{r\in \{s\} \cup {\cal S}_s } \!\!\!  \alpha_{s,r} \ist D_{r,k}^{\text{GM}[i-1]}(\mathbf{x}) \ist.
\label{eq:GM-PHD_cons-fused_2}
\vspace{-.5mm}
\end{equation} 
Hence, we have emulated the 
GM-PHD average consensus iteration \eqref{eq:D_averaging} by operating at the level of the GM parameters. Finally,
a merging step reduces $\mathcal{G}_{s,k}^{[i-1],\cup}$ to a smaller GM parameter set 
\[
\mathcal{G}_{s,k}^{[i]} \rmv\triangleq \big\{ \big( \Omega_{s,k,\ell}^{[i]}, \bm{\mu}_{s,k,\ell}^{[i]},\bm{\Sigma}_{s,k,\ell}^{[i]} \big) \big\}_{\ell \in L_{s,k}^{[i]}}\!.
\vspace{-1mm}
\]
The GM-PHD corresponding to $\mathcal{G}_{s,k}^{[i]}$, i.e., 
\be
D_{s,k}^{\text{GM}[i]}(\mathbf{x}) \triangleq\! \sum_{\ell \in L_{s,k}^{[i]}} \!\! \Omega_{s,k,\ell}^{[i]} 
  \,\mathcal{N}\big(\mathbf{x};\bm{\mu}_{s,k,\ell}^{[i]}, \bm{\Sigma}_{s,k,\ell}^{[i]} \big) \ist,
\label{eq:GM-PHD_cons-fused_2_merged}
\vspace{-.5mm}
\ee
approximates $D_{s,k}^{\text{GM}[i-1],\cup}(\mathbf{x})$ in \eqref{eq:GM-PHD_cons-fused_2}. The recursion $\mathcal{G}_{s,k}^{[i-1]} \!\to \mathcal{G}_{s,k}^{[i]}$ 
described above is initialized with $\mathcal{G}_{s,k}^{[0]} \!=\rmv \mathcal{G}_{s,k}$.

Thus, after 
$I$ iterations, we have converted the original local GM parameter set $\mathcal{G}_{s,k}$ into a fused GM parameter set $\mathcal{G}_{s,k}^{[I]}$
that approximately emulates $I$ average consensus iterations \eqref{eq:D_averaging}.
The choice of 
$I$ will be discussed in Section \ref{sec:summ-parallel}.
In conclusion, we have developed an approximate implementation of the GM-PHD average consensus scheme \eqref{eq:D_averaging} that operates at the level of the GM parameters. 
Note that here---in contrast to the distributed flooding scheme discussed in Section \ref{sec:P2GM_flood}---the 
iterated GM parameter sets $\mathcal{G}_{s,k}^{[i]}$ do not systematically grow
with progressing iteration $i$. Furthermore, our experimental results reported in Section \ref{sec:simulation} suggest that the proposed GM average consensus scheme
with GC merging can outperform the GM flooding scheme in terms of tracking accuracy.

%





\section{IS Method for GM--Particles Conversion} 
\label{sec:IS}

The dissemination/fusion schemes discussed in the previous section effectively provide each sensor $s$ with a fused GM-PHD $D_{s,k}^{\text{GM}[I]}(\mathbf{x})$,
which is given by \eqref{eq:D_flooding} if the GM flooding scheme of Section \ref{sec:P2GM_flood} is used and by \eqref{eq:GM-PHD_cons-fused_2_merged}
(with $i$ replaced by $I$) if the GM average consensus
scheme of Section \ref{sec:P2GM_merg} is used. (We say ``effectively'' because $D_{s,k}^{\text{GM}[I]}(\mathbf{x})$ is not actually calculated.)
In what follows, we will denote by 
\begin{equation} \label{eq:fused_GM}
\mathcal{G}_{s,k}^{[I]} \triangleq \big\{ \big( \Omega_{s,k,\ell}^{[I]}, \bm{\mu}_{s,k,\ell}^{[I]},\bm{\Sigma}_{s,k,\ell}^{[I]} \big) \big\}_{\ell \in L_{s,k}^{[I]}}
\vspace{-.5mm}
\end{equation}
the set of GM parameters involved in
$D_{s,k}^{\text{GM}[I]}(\mathbf{x})$, i.e., we have
\be 
\label{eq:consensus_D_fused} 
D_{s,k}^{\text{GM}[I]}(\mathbf{x}) =\! \sum_{\ell \in L_{s,k}^{[I]}} \!\! \Omega_{s,k,\ell}^{[I]} \,\mathcal{N}\big(\mathbf{x};\bm{\mu}_{s,k,\ell}^{[I]}, \bm{\Sigma}_{s,k,\ell}^{[I]} \big) \ist.
\vspace{-1.3mm}
\ee
Here,
in the case of GM flooding, $\mathcal{G}_{s,k}^{[I]}$ is 
obtained from $\mathcal{G}_{s,k}^{\text{F}[I]}$ in \eqref{eq:GM_flooding}
by scaling all the weights in $\mathcal{G}_{s,k}^{\text{F}[I]}$ with the factor $1/\big| {\cal S}_s^{[I]} \big|$; this 
accounts for the factor $1/\big| {\cal S}_s^{[I]} \big|$ in \eqref{eq:D_flooding}.


In order to use
the fused GM-PHD $D_{s,k}^{\text{GM}[I]}(\mathbf{x})$ in the local particle-PHD filter at sensor $s$, it is necessary to find a particle representation of 
$D_{s,k}^{\text{GM}[I]}(\mathbf{x})$.
The standard method is
to sample directly from $D_{s,k}^{\text{GM}[I]}(\mathbf{x})$.
However, we here propose a method\linebreak 
based on the importance sampling (IS) principle \cite[Ch. 3.3]{Robert05}, which will be seen in Section \ref{sec:summ-parallel} 
to enable a 
parallelization of filtering and fusion operations. 
We start by recalling from Section \ref{sec:phd} that the local PHD filter 
propagates over time $k$ a weighted particle set $\xi_{s,k} = \big\{ \big( \mathbf{x}_{s,k}^{(j)} \ist, w_{s,k}^{(j)}\big) \big\}_{j=1}^{J_{s,k}}$  providing an approximate 
representation of 
$\hat{D}_{s,k}(\mathbf{x}|Z_{s,1:k})$. 
Let us now
consider an alternative particle representation of $\hat{D}_{s,k}(\mathbf{x}|Z_{s,1:k})$
using a uniformly weighted particle set $\big\{ \big( \tilde{\mathbf{x}}_{s,k}^{(j)} \ist, c_{s,k} \big) \big\}_{j=1}^{\tilde{J}_{s,k}}$. Here, 
the number of uniformly weighted particles is chosen \nolinebreak 
as
\be
\tilde{J}_{s,k} = \mathrm{round} \{ N_{\textrm{p}} \ist W_{s,k} \} \ist,
\label{eq:Jsk_round}
\vspace{-.5mm}
\ee
where $N_\textrm{p} \!\in\! \mathbb{N}$ is a parameter specifying the number of particles assigned to each potential target, as discussed 
in \cite[Sec.~III.C]{Vo05}, and, as before (see \eqref{eq:W_k:z_def}), $W_{s,k}$ is the sum of the original weights $w_{s,k}^{(j)}$.
Furthermore, the weight $c_{s,k}$---identical for all $j$---is 
\[
c_{s,k} = \frac{W_{s,k}}{\tilde{J}_{s,k}} \,.
\vspace{-1mm}
\] 

The new particles $\tilde{\mathbf{x}}_{k}^{(j)}$ are obtained from the original weighted particle set $\xi_{s,k}$ 
through
resampling, which means that particles with large weights are replicated whereas those with small weights are removed
\cite{Li15SPM}. 
As such, each resampled particle $\tilde{\mathbf{x}}_{s,k}^{(j)}$ equals one of the original particles, $\mathbf{x}_{s,k}^{(j')}$, where 
$j'$ is uniquely determined by $j$. 
Note that some of the $\tilde{\mathbf{x}}_{s,k}^{(j)}$ are identical due to the replication.
Let $N_{s,k}^{(j')}$ denote the number of\linebreak 
times 
particle $\mathbf{x}_{s,k}^{(j')}$ is resampled (replicated). To ensure unbiased resampling, we require that the expectation of $N_{s,k}^{(j')}$ given 
$\xi_{s,k} \rmv=\rmv \big\{ \big( \mathbf{x}_{s,k}^{(j)} \ist, w_{s,k}^{(j)}\big) \big\}_{j=1}^{J_{s,k}}$
is $N_{\textrm{p}}$ times
$w_{s,k}^{(j')}$ \cite{Li15SPM}, i.e.,
\be
\mathrm{E} \big[N_{s,k}^{(j')} \big| \xi_{s,k} \big] = N_{\textrm{p}} \ist w_{s,k}^{(j')}.
\label{eq:E_resampling}
\ee 
As verified in Appendix \ref{appendix:A}, this can be achieved approximately by choosing a new particle $\tilde{\mathbf{x}}_{s,k}^{(j)}$ equal to $\mathbf{x}_{s,k}^{(j')}$ with 
\vspace{-1mm}
probability
\begin{equation} 
\label{eq:ResamplingProposal}
P_{j'} \triangleq\ist \mathrm{Pr}\big[\tilde{\mathbf{x}}_{s,k}^{(j)} \!=\rmv \mathbf{x}_{s,k}^{(j')} \big| \xi_{s,k} \big]
\rmv= \frac{w_{s,k}^{(j')}}{W_{s,k}} \ist.
\vspace{-1.5mm}
\end{equation}


The resampled particle set $\big\{ \big( \tilde{\mathbf{x}}_{s,k}^{(j)} \ist, c_{s,k} \!=\rmv W_{s,k}/\tilde{J}_{s,k}\big) \big\}_{j=1}^{\tilde{J}_{s,k}}$ represents 
$\hat{D}_{s,k}(\mathbf{x}|Z_{s,1:k})$. However, based on the IS principle \cite[Ch. 3.3]{Robert05}, we can also use 
$\big\{ \tilde{\mathbf{x}}_{s,k}^{(j)} \big\}_{j=1}^{\tilde{J}_{s,k}}$ to represent the fused 
GM-PHD\footnote{This 
representation can be expected to be accurate only if the effective support of $D_{s,k}^{\text{GM}[I]}(\mathbf{x})$ is 
contained in that of $\hat{D}_{s,k}(\mathbf{x}|Z_{s,1:k})$. This condition is satisfied for all $s$ if the fields of view of all sensors are effectively 
equal. In the opposite case, one has to expect a performance loss compared to the standard method of sampling directly from 
$D_{s,k}^{\text{GM}[I]}(\mathbf{x})$.} 
$D_{s,k}^{\text{GM}[I]}(\mathbf{x})$ in \eqref{eq:consensus_D_fused}, if only the weight associated with $\tilde{\mathbf{x}}_{s,k}^{(j)} = \mathbf{x}_{s,k}^{(j')}$ is chosen as
\be
\tilde{w}_{s,k}^{(j)} 
\ist=\ist \frac{ D_{s,k}^{\text{GM}[I]}\big(\tilde{\mathbf{x}}_{s,k}^{(j)} 
\big) }{ P_{j'} } 
  \ist=\ist \frac{W_{s,k} \ist D_{s,k}^{\text{GM}[I]}\big(\tilde{\mathbf{x}}_{s,k}^{(j)} \big) }{w_{s,k}^{(j')} } \ist, 
\label{eq:W_ratio_imf}
\vspace{-2mm}
\ee
where, from \eqref{eq:consensus_D_fused},
\be
D_{s,k}^{\text{GM}[I]}\big( \tilde{\mathbf{x}}_{s,k}^{(j)} \big) =\! \sum_{\ell \in L_{s,k}^{[I]}} \!\! \Omega_{s,k,\ell}^{[I]} 
  \,\mathcal{N}\big(\tilde{\mathbf{x}}_{s,k}^{(j)};\bm{\mu}_{s,k,\ell}^{[I]}, \bm{\Sigma}_{s,k,\ell}^{[I]} \big) \ist, 
\label{eq:D_C_j} 
\vspace{-2mm}
\ee
with $\tilde{\mathbf{x}}_{s,k}^{(j)} \!=\! \mathbf{x}_{s,k}^{(j')}$. Hereafter, we use 
$\big\{ \big( \tilde{\mathbf{x}}_{s,k}^{(j)} \ist, \tilde{w}_{s,k}^{(j)} \big) \big\}_{j=1}^{\tilde{J}_{s,k}}$ to represent $D_{s,k}^{\text{GM}[I]}(\mathbf{x})$. 
The particle set conversion $\xi_{s,k} \rmv\to \big\{ \big( \tilde{\mathbf{x}}_{s,k}^{(j)} \ist, \tilde{w}_{s, k}^{(j)} \big) \big\}_{j=1}^{\tilde{J}_{s,k}}$
developed above constitutes a particle implementation of the PHD fusion conversion $\hat{D}_{s,k}(\mathbf{x}|Z_{s,1:k}) \rmv\to D_{s,k}^{\text{GM}[I]}(\mathbf{x})$.



\section{Cardinality Averaging and State Estimation}
\label{sec:rest}

\vspace{.5mm}

Next, we discuss two final stages of our distributed PHD filtering method.

\vspace{-1mm}

\subsection{AA-based Cardinality Averaging} 
\label{sec:CC-AA}

By \eqref{eq:W_k:j}, $W_{s,k} \!=\! \sum_{\mathbf{z} \in Z_{s,k}^{0}} \!\! \Omega_{s,k}(\mathbf{z})$ (see \eqref{eq:W_k_Omega})
provides \nolinebreak 
an \nolinebreak 
estimate \nolinebreak 
of the cardinality
$N_k \rmv=\rmv | X_k |$. However, in our particles--GM conversion method 
(see Section \ref{sec:P2GM}), $Z_{s,k}^{0}$ was replaced by the subset 
$Z_{s,k}^\text{S}$, and consequently $\sum_{\mathbf{z} \in Z_{s,k}^{0}} \!\! \Omega_{s,k}(\mathbf{z})$ is 
replaced by $\sum_{\mathbf{z} \in Z_{s,k}^\text{S}} \!\! \Omega_{s,k}(\mathbf{z}) \le \sum_{\mathbf{z} \in Z_{s,k}^{0}} \!\! \Omega_{s,k}(\mathbf{z})$.
This implies that the\linebreak 
fused GM-PHD $D_{s,k}^{\text{GM}[I]}(\mathbf{x})$ in \eqref{eq:consensus_D_fused} and the associated weights $\tilde{w}_{s,k}^{(j)}$ 
in \eqref{eq:W_ratio_imf} will both underestimate the cardinality $N_k$, in the sense that, typically, 
$\int_{\mathbb{R}^d} \rmv D_{s,k}^{\text{GM}[I]}(\mathbf{x}) \ist \mathrm{d}\mathbf{x} \rmv<\! N_k$ and
$\sum _{j=1}^{\tilde{J}_{s,k}} \rmv \tilde{w}_{s, k}^{(j)}$\linebreak 
$<\rmv N_k$.

This ``cardinality bias'' can be compensated by a suitable
scaling of the weights $\tilde{w}_{s, k}^{(j)}$. In our method (see Steps 
4 and 5 in Section \ref{sec:motivation-outline}), 
following \cite{Li19CC}, this 
scaling is based on the original---``correct''--- local cardinality estimates $W_{s,k}$, which are averaged over all sensors $s$ to smooth out 
sensor-specific errors. That is, we attempt to calculate the AA of all the local cardinality estimates, 
$\overline{W}_{\rmv k} \triangleq \sum _{s=1}^S \rmv W_{s,k}/S$,
and use the result to 
scale the 
$\tilde{w}_{s, k}^{(j)}$. 
Note that $\overline{W}_{\rmv k} \rmv=\rmv \int_{\mathbb{R}^d} \rmv \hat{D}_{k}(\mathbf{x}|Z_{1:S,1:k}) \ist \mathrm{d}\mathbf{x}$ with
$\hat{D}_{k}(\mathbf{x}|Z_{1:S,1:k}) = \sum_{s=1}^S \hat{D}_{s,k}(\mathbf{x}|Z_{s,1:k})/S$ 
\vspace{.6mm} 
as defined in \eqref{eq:av-local-PHDs}, which means that 
$\overline{W}_{\rmv k}$ is the cardinality estimate based on the AA of all the local PHDs $\hat{D}_{s,k}(\mathbf{x}|Z_{s,1:k})$.

For a distributed approximate calculation of $\overline{W}_{\rmv k}$, 
we can use 
flooding or average consensus on the $W_{s,k}$
(cf.\ Section \ref{sec:GM-cons}) \cite{Li19CC}. 
Let $W_{\! s,k}^{[I]}$ be
the approximation of $\overline{W}_{\rmv k}$ obtained after $I$ flooding or average consensus iterations.
Then, the 
weights $\tilde{w}_{s, k}^{(j)}$
are scaled 
\vspace{-.8mm}
as
\begin{equation}\label{eq:CC_W_scaling_0}
\bar{w}_{s, k}^{(j)} = \beta_{s,k} \ist \tilde{w}_{s, k}^{(j)} \ist, \quad j \rmv=\rmv 1,\ldots,\tilde{J}_{s,k} \ist,
\end{equation}
where, as derived in \cite{Li19CC},
\vspace{-1mm}
\be
\label{eq:beta}
\beta_{s,k} = \frac{ W_{\! s,k}^{[I]} }{ \sum _{j=1}^{\tilde{J}_{s,k}} \tilde{w}_{s, k}^{(j)} } \ist.
\ee 
We then use $\big\{ \big( \tilde{\mathbf{x}}_{s,k}^{(j)} \ist, \bar{w}_{s, k}^{(j)} \big) \big\}_{j=1}^{\tilde{J}_{s,k}}$ as the final particle representation of the fused PHD
$D_{s,k}^{\text{GM}[I]}(\mathbf{x})$. In the local PHD filter at sensor $s$, $\big\{ \big( \tilde{\mathbf{x}}_{s,k}^{(j)} \ist, \bar{w}_{s, k}^{(j)} \big) \big\}_{j=1}^{\tilde{J}_{s,k}}$
replaces the original particle representation $\xi_{s,k}$, 
i.e., it is used instead of $\xi_{s,k}$ in the next prediction step. 
We note that an accurate cardinality estimate
is also crucial for target state estimation,
as explained next.

\vspace{-1mm}

\subsection{Target State Estimation}
\label{sec:EstimateExtraction}

At each sensor $s$ and time $k$, estimates of the target states $\mathbf{x}_k^{(\nu)}$
are calculated
as follows. First, 
\pagebreak 
an estimate of the number of targets
is formed as $\hat{N}_{s,k} \rmv\triangleq \mathrm{round}\big\{ W_{\! s,k}^{[I]} \big\}$, where $W_{\! s,k}^{[I]}$ is the result of the cardinality averaging scheme discussed above.
Then, the means of the $\hat{N}_{s,k}$ GCs with the largest weights $\Omega_{s,k,\ell}^{[I]}$ are used as estimates of the target 
states.\footnote{An 
alternative method is to group all the GC means 
into $\hat{N}_{s,k}$ clusters and use the weighted average
of the means of each cluster as a state estimate. However, this method is more complex and, moreover,
did not perform better in our 
simulations.} 
Note that this target state estimation
operation is performed 
locally at sensor $s$.

\begin{algorithm}[t!] \label{alg1}
\small
\caption{Proposed distributed \vspace{-.3mm}
particle-PHD filter algo\-rithm---operations performed 
at sensor $s$ during time step $k$}
\vspace{1.5mm}
\textbf{Input:\,}
Previous particle set $\big\{ \big( \mathbf{x}_{s,k-1}^{(j)} \ist, w_{s, k-1}^{(j)} \big) \big\}_{j=1}^{J_{s,k-1}}$;
measurement set $Z_{s, k}$; number of newborn particles $L_{s,k}$. 


\vspace{.7mm}

\textbf{Output:\,} 
New particle set $\big\{ \big( \tilde{\mathbf{x}}_{s,k}^{(j)} \ist, \bar{w}_{s, k}^{(j)} \big) \big\}_{j=1}^{\tilde{J}_{s,k}}$
(this particle set will be used as the input---see above---at
the next time step $k \rmv+\! 1$);
target state 
estimates $\hat{\mathbf{x}}_{s,k}^{(\nu)}$, $\nu = 1,\ldots, \hat{N}_{s,k}$.


\vspace{1.6mm}

\textbf{Operations:} 

\vspace{1.5mm}

\hspace{1.5mm} \emph{\textbf{Local filtering}}

\vspace{.5mm}

\begin{enumerate}

\item

For $j \rmv=\rmv 1,\ldots,J_{s,k}$, with $J_{s,k} \rmv=\rmv J_{s,k-1} \!+\! L_{s,k}$, draw particles\linebreak 
$\mathbf{x}_{s,k}^{(j)}$ from proposal pdf $q_{s,k}\big(\mathbf{x} ; \mathbf{x}_{s,k-1}^{(j)},Z_{s,k}\big)$ 
\vspace{.4mm} 
(if $j \rmv\in\rmv \{1,\ldots,$\linebreak 
$J_{s,k-1}\}$) or $p_{s,k}(\mathbf{x} ; Z_{s,k})$ (if $j \rmv\in\rmv \{J_{s,k-1} \!+\! 1,\ldots,J_{s,k}\}$). 

\vspace{1mm}

\item 
Evaluate
$\rmv f_{k}\big(\mathbf{x}_{s,k}^{(j)} \big| \mathbf{x}_{s,k-1}^{(j)}\big)\rmv$ and $q_{s,k}\big(\mathbf{x}_{s,k}^{(j)} ; \mathbf{x}_{s,k-1}^{(j)},Z_{s,k}\big)$ for 
$j \!=$\linebreak 
$1, \ldots, J_{s,k-1}$;
$\gamma_k\big(\mathbf{x}_{s,k}^{(j)}\big)$ and $p_{s,k}\big(\mathbf{x}_{s,k}^{(j)}; Z_{s,k}\big) $ for 
$j \rmv=\rmv J_{s,k-1}$\linebreak 
$+\ist 1,\ldots,J_{s,k}$;
$p_{s,k}^\text{D}\big(\mathbf{x}_{s, k}^{(j)}\big)$ for $j \rmv=\rmv 1, \ldots, J_{s,k}$;
$g_{s,k}\big(\mathbf{z} \big| \mathbf{x}_{s,k}^{(j)}\big)$ for $\mathbf{z} \!\in\! Z_{s, k}$ and $j \rmv=\rmv 1, \ldots, J_{s,k}$; and
$\kappa_{s,k}(\mathbf{z})$ for $\mathbf{z} \!\in\! Z_{s,k}$. 

\vspace{1mm}

\item 
Calculate $w_{s,k|k-1}^{(j)}$ for $j \rmv=\rmv 1, \ldots, J_{s,k}$ using \eqref{eq:w_k|k-1_j}.
 
\vspace{1mm}

\item 
Calculate $\omega_{s,k}^{(j)}(\mathbf{z})$ for 
$\mathbf{z} \!\in\! Z_{s,k}^{0}$ and $j \rmv=\rmv 1, \ldots, J_{s,k}$ using \eqref{eq:w_decomposition}.

\vspace{1mm}

\item 
Calculate $w_{s,k}^{(j)}$ for $j \rmv=\rmv 1, \ldots, J_{s,k}$ using \eqref{eq:w_k(j)}.

\vspace{1mm}

\item
Calculate $W_{s,k}$ according to \eqref{eq:W_k:z_def}.

\vspace{1mm}

\item
Resample $\big\{ \big( \mathbf{x}_{s,k}^{(j)} \ist, w_{s,k}^{(j)}\big) \big\}_{j=1}^{J_{s,k}}$ to obtain a uniformly weighted particle set 
$\big\{ \big( \tilde{\mathbf{x}}_{s,k}^{(j)} \ist, 
c_{s,k} \big) \big\}_{j=1}^{\tilde{J}_{s,k}}$, where $c_{s,k} = W_{s,k} \ist /\tilde{J}_{s,k}$ with $\tilde{J}_{s,k} \!=\rmv \mathrm{round} \{ N_{\textrm{p}} \ist W_{s,k} \}$. For 
$j \rmv=\rmv 1,\ldots,\tilde{J}_{s,k}$, store the weight $w_{s,k}^{(j')}$ of the particle $\mathbf{x}_{s, k}^{(j')}$ associated with $\tilde{\mathbf{x}}_{s,k}^{(j)}$.

\vspace{2mm}

\hspace{-7mm} \emph{\textbf{Fusion}}

\vspace{1mm}

\item 
Calculate $\Omega_{s,k}(\mathbf{z})$ for 
$\mathbf{z} \rmv\in\! Z_{s,k}^{0}$ according to \eqref{eq:W_k(z)}. 

\vspace{1mm}

\item 
Determine the subset of significant measurements, $Z_{s,k}^\text{S} \!\subseteq\! Z_{s,k}^{0}$, as the set of those $\mathbf{z} \!\in\! Z_{s,k}^{0}$ for which $\Omega_{s,k}(\mathbf{z}) > T_\Omega$.

\vspace{1mm}

\item
For 
$\mathbf{z} \!\in\! Z_{s,k}^\text{S}$, determine 
$\bm{\mu}_{s,k}(\mathbf{z})$ and $\bm{\Sigma}_{s,k}(\mathbf{z})$ according to \eqref{eq:GC_mean} and \eqref{eq:GC_P}, respectively. 

\vspace{1mm}

\item 
Calculate the fused GM parameter set $\mathcal{G}_{s,k}^{[I]} \rmv=\rmv \big\{ \big( \Omega_{s,k,\ell}^{[I]}, \bm{\mu}_{s,k,\ell}^{[I]},$\linebreak 
$\bm{\Sigma}_{s,k,\ell}^{[I]} \big) \big\}_{\ell \in L_{s,k}^{[I]}}$ (cf.\ \eqref{eq:fused_GM})
by performing $I$ iterations of a distributed dissemination/fusion scheme as described in Section \ref{sec:GM-cons}. This requires broadcasting data to 
sensors $r \!\in\! {\cal S}_s$.

\vspace{1mm}

\item 
Calculate the fused cardinality estimate $W_{\! s,k}^{[I]}$ by means of
distributed cardinality averaging as described in Section \ref{sec:CC-AA}.
This requires broadcasting data to 
sensors $r \!\in\! {\cal S}_s$.

\vspace{1.1mm}

\item
Calculate $D_{s,k}^{\text{GM}[I]}\big(\tilde{\mathbf{x}}_{s,k}^{(j)}\big)$ for $j \rmv=\rmv 1,\ldots,\tilde{J}_{s,k}$ using \eqref{eq:D_C_j}.

\vspace{1mm}

\item
Calculate $\tilde{w}_{s, k}^{(j)}$ for $j=1,\ldots,\tilde{J}_{s,k}$ using \eqref{eq:W_ratio_imf}.

\vspace{1mm}

\item
Calculate $\bar{w}_{s, k}^{(j)}$ for $j \rmv=\rmv 1,\ldots,\tilde{J}_{s,k}$ using \eqref{eq:CC_W_scaling_0} and \eqref{eq:beta}.

\vspace{2.1mm}

\hspace{-6.9mm} \emph{\textbf{Target state estimation}}

\vspace{1mm}


\item
Calculate an estimate of the number of targets as $\hat{N}_{s,k} \rmv= \mathrm{round}\big\{ W_{\! s,k}^{[I]} \big\}$. 

\vspace{.7mm}
 
\item
Take the target state estimates $\hat{\mathbf{x}}_{s,k}^{(\nu)}$, $\nu = 1,\ldots, \hat{N}_{s,k}$ to be the means of the $\hat{N}_{s,k}$ GCs with the largest weights $\Omega_{s,k,\ell}^{[I]}$.

\end{enumerate}

\end{algorithm}


\section{Algorithm Summary, Parallelization,
Communication Cost}
\label{sec:summ_parallel_comm}

\vspace{1mm}

\subsection{Algorithm Summary and Parallelization} 
\label{sec:summ-parallel}

\begin{figure}
\vspace{-1mm}
\centering
\includegraphics[width=9 cm]{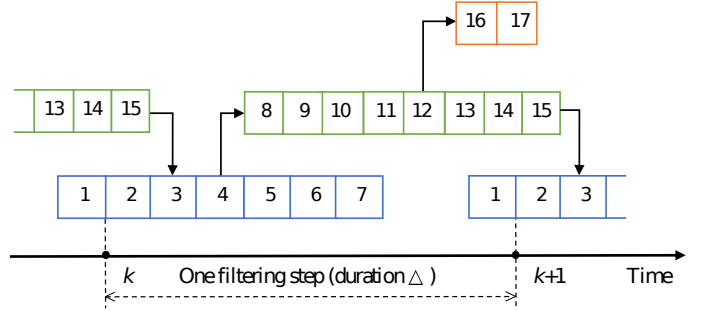}
\vspace{-7mm}
\caption{
Parallelization of the operations of Algorithm 1. The numbers shown equal the operation numbers used in Algorithm 1.
We note that the temporal duration of operations 11 and 12 
is proportional to the number $I$ of dissemination/fusion 
iterations, which is upper bounded by $I_\text{max}$ in \eqref{eq:T_bound}.}
\label{fig:diagram}
\vspace{-2mm}
\end{figure}

A summary of the proposed distributed 
PHD filter algorithm is given in Algorithm 1. Two noteworthy aspects 
are that (i) the\linebreak 
filtering operations 1 and 2 do not require or change the previous particle weights,
and (ii) the fusion-related operations 8--15 do not change the current particles.
As a consequence, the filtering operations 1 and 2 
for time $k \rmv+\! 1$ can be carried out in parallel (simultaneously) with the fusion-related operations 
8--15 for time $k$. More specifically, operations 1 and 2 for time\linebreak 
$k \rmv+\! 1$ can be carried out as soon as operation 7 
for time $k$ is done;
they do not need to wait for the results of operations 8--17. Also, operations 8--10 for time $k$ can be performed 
in\linebreak 
parallel with operations 5--7 
for time $k$. 
In summary, the filtering operations 1 and 2 for time $k+1$ and the filtering operations 5--7 for time $k$
can be performed in parallel with the fusion-related operations 8--15 for time $k$.
Since operation 2 (including calculation of 
$g_{s,k}\big(\mathbf{z} \big|\mathbf{x}_{s,k}^{(j)}\big)$) and operation 7 (resampling) are the most computationally 
intensive filtering operations, a large degree of parallelization is possible. A timing diagram illustrating the scheduling and parallelization of the various operations is given in 
Fig.\ \ref{fig:diagram}. 

This parallelization, which is enabled by
our IS method for GM--particles conversion, is an important advantage of the proposed distributed PHD filter algorithm.
Indeed, with most other distributed PHD filtering algorithms,
the filtering operations can only be scheduled 
before or after the dissemination/fusion operations. Because the time duration $\Delta$ of one 
filtering step
(corresponding to one time step $k \to k+1$) is limited by the time between two sensing scans, this serial schedule implies a strong limitation of the number $I$ of dissemination/fusion 
iterations that can be carried out in each filtering step.
More specifically, for any distributed filtering algorithm, the maximum possible value of $I$ is 
\be 
\label{eq:T_bound}
I_\text{max} = \bigg\lfloor \frac{\Delta \rmv-\rmv t_\text{filt} \rmv-\rmv t_\text{inter} }{\tau} \bigg\rfloor \ist.
\ee
Here, $t_\text{filt}$ is the total time duration of all the filtering \nolinebreak 
operations that cannot be carried out in parallel with the dissemination/fusion iterations;
$t_\text{inter}$ is the time required \nolinebreak 
by  \nolinebreak 
operations interfacing the dissemination/fusion scheme with the local filtering (preparing data to be communicated, 
inserting the communicated data into the local filter, etc.), which have to be\linebreak 
performed before and/or after the dissemination/fusion iterations;
and $\tau$ is the time duration of one dissemination/fusion iteration. 
With
our algorithm, operations 3 and 4 contribute to\linebreak 
$t_\text{filt}$ and operations 8--10 and 13--15 contribute to $t_\text{inter}$. Here, $t_\text{inter}$ is comparable to
most other algorithms but $t_\text{filt}$ is significantly smaller. 
In fact, for most other algorithms, $t_\text{filt}$ is the\linebreak 
total duration of all the filtering operations (cf.\
our operations 1--7), which includes also the computationally 
intensive likelihood calculation (cf.\ operation 2) and, for,
a particle-based implementation, also resampling (cf.\ operation 7). Thus, it follows from \eqref{eq:T_bound} that for our
algorithm, $I_\text{max}$ is significantly larger than for 
the other algorithms. This is an important advantage, as more dissemination/fusion iterations 
usually imply\linebreak 
a better 
estimation accuracy. 


\vspace{-1mm}


\subsection{Communication Cost} 
\label{sec:comm_cost}

In one dissemination/fusion 
iteration of the proposed \nolinebreak 
distributed \nolinebreak 
PHD filter, each sensor $s$ broadcasts to its neighbors
a certain number of GC parameter sets, where each set consists of a weight, a $d$-dimensional mean vector, and a $d\times d$ symmetric covariance matrix.
Thus, for each GC, $n_\textrm{GC} \triangleq 1 + d + \frac{d(d+1)}{2}$ real values are
broadcast by sensor $s$. In addition, sensor $s$ broadcasts one cardinality estimate, which is a single real value.
Let $n_{s,k}^{[i]}$ denote the number of GCs contained in the GM of sensor $s$ in dissemination/fusion iteration $i$, 
before the fusion with the neighboring sensors is performed. Then the total number of real values broadcast by sensor $s$ in one 
dissemination/fusion 
iteration is
\begin{equation} 
\label{eq:flooding_communication_cost}
N_{s,k}^\text{com\ist[i]} = n_{s,k}^{[i]} n_\textrm{GC} + 1 = n_{s,k}^{[i]} \bigg( 1 + d + \frac{d(d+1)}{2} \bigg) + 1 .   
\end{equation} 
Note that $N_{s,k}^\text{com\ist[i]}$
grows linearly with the number of GCs, $n_{s,k}^{[i]}$, and quadratically with the dimension 
of the target states, $d$, and it does not depend on the number of sensors, $S$. The last fact implies that the total communication cost
for the entire network grows linearly with the network size $S$.

While expression \eqref{eq:flooding_communication_cost} holds for both the GM flooding scheme of Section \ref{sec:P2GM_flood} and the GM average consensus 
scheme of Section \ref{sec:P2GM_merg}, the communication costs of the two schemes are actually very different. In the case of the GM flooding scheme,
the number of GCs broadcast is $n_{s,k}^{[i]} \!=\! \big| \mathcal{G}_{s,k}^{\text{F}[i-1]} \big|$, which systematically grows with the iteration index $i$ 
according to \eqref{eq:GM_flooding_i} or equivalently \eqref{eq:GM_flooding_i_2}. In the case of the GM average consensus scheme, we have
$n_{s,k}^{[i]} \!=\! \big| \mathcal{G}_{s,k}^{[i-1]} \big|$, which, according to Section \ref{sec:P2GM_merg}, does not systematically grow with $i$ because in each iteration a GC merging step is carried out.
A quantitative characterization of $\big| \mathcal{G}_{s,k}^{[i-1]} \big|$ is difficult because the reduction of the number of GCs due to merging
is larger if the GCs are closer to each other.

\section{Simulation Study}
\label{sec:simulation}


\subsection{Simulation Setup} 
\label{sec:SimulationSet}

\subsubsection{Targets and Sensors} 
\label{sec:SimulationSet_targets}
\begin{figure}
\vspace{-2mm}
\centering
\includegraphics[width=7 cm]{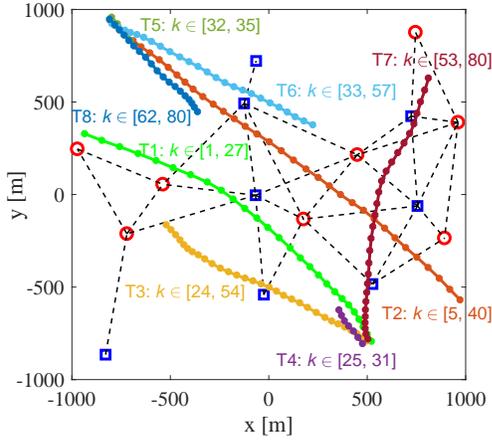}
\vspace{-1.5mm}
\caption{ROI, sensor network, and target trajectories. Blue squares and red
circles indicate the positions of the linear and nonlinear sensors, respectively, black dashed lines indicate the communication links between
neighboring sensors, and colored lines with dots indicate the target
trajectories (with starting and ending times noted).}
\label{fig:scene1}
\vspace{-1mm}
\end{figure}

We simulated six targets that move in a square two-dimen\-sional (2-D) region of interest (ROI) given by 
$[-1000\ist\text{m}, 1000\ist\text{m}]\times [-1000\ist\text{m}, 1000\ist\text{m}]$.
The sensor network---consisting of 16 sensors---and
the target trajectories
are depicted in Fig.\ \ref{fig:scene1}.
The target states consist of 2-D position and 2-D velocity, i.e., $\mathbf{x}_k=[x_k \; \dot{x}_k \; y_k \; \dot{y}_k]^\text{T}\rmv$.
The target survival probability is $p_{k}^\text{S}(\mathbf{x}_{k-1}) \!=\! 0.98$.
The states of the surviving targets evolve independently according to a nearly constant velocity model, i.e.,
$\mathbf{x}_k \rmv=\rmv \mathbf{F}\mathbf{x}_{k-1} \rmv+ \mathbf{G}\mathbf{u}_{k}$,
where $\mathbf{F} \!\in\! \mathbb{R}^{4\times 4}\rmv$ and\linebreak 
$\mathbf{G} \!\in\! \mathbb{R}^{4\times 2}\rmv$ are as given in
\cite[Eq.\ (14)]{Li03} with sampling period $\Delta \!=\! 1\ist$s and $\mathbf{u}_{k}$ is an independent and identically distributed (iid),
zero-mean, Gaussian system process with standard deviation $5\ist$m/s$^2$.
The birth intensity function is 
$\gamma_k(\mathbf{x}_k)= 0.05 \cdot \mathcal{N}(\mathbf{x}_k;\mathbf{m}_1, \mathbf{Q}) + 0.05 \cdot \mathcal{N}(\mathbf{x}_k;\mathbf{m}_2, \mathbf{Q})$,
where 
$\mathbf{m}_1 \!= [{500}\ist\text{m} \; {-20}\ist\text{m}/\text{s}\; {-800}\ist\text{m} \; 30\ist\text{m}/\text{s}]^\text{T}\rmv$, 
$\mathbf{m}_2 \!=\rmv [{-800}\ist\text{m} \; 30\ist\text{m}/\text{s}$\linebreak 
${950}\ist\text{m} \; {-30}\ist\text{m}/\text{s}]^\text{T}\rmv$, 
and $\mathbf{Q} \rmv=\rmv\text{diag}\{ 400\ist\text{m}^2\rmv, 100\ist\text{m}^2/\text{s}^2\rmv, 400\ist\text{m}^2\rmv,$\linebreak 
$100\text{m}^2/\text{s}^2 \}$. 

Eight of the 16 sensors acquire noisy position measurements within the ROI
with a fixed detection probability $p_{s,k}^\text{D}(\mathbf{x}_k) \rmv= 0.9$. 
For these ``linear'' sensors, the measurement model 
\vspace{-.2mm}
is
\[
\mathbf{z}_{s,k} = [x_k \; y_k]^{\textrm{T}} + \big[v_{s,k}^{(1)} \;\ist v_{s,k}^{(2)}\big]^{\textrm{T}} \rmv,
\vspace{-1mm}
\]
where $v_{s,k}^{(1)}$ and $v_{s,k}^{(2)}$ are iid zero-mean Gaussian with standard deviation $20\ist$m$^2$.
The other eight sensors are ``nonlinear'' sensors that acquire noisy range and bearing measurements with 
detection probability given by \cite{Vo09}
\[
p_{s,k}^\text{D}(\mathbf{x}_k) = 0.95 \cdot \frac{ \mathcal{N}\big( \bm{\mu}_\mathrm{D}(\mathbf{x}_k); \mathbf{0},6000^2 \mathbf{I}_2\big)}{
  \mathcal{N}(\mathbf{0};\mathbf{0},6000^2\mathbf{I}_2)} \ist.
\vspace{-.5mm}
\]
Here, 
$\bm{\mu}_\mathrm{D}(\mathbf{x}_k) \rmv\triangleq\rmv \big[ \ist | x_k \!-\! x^{(s)} | \;\ist | y_k \!-\! y^{(s)} | \ist\big]^\text{T}\!$, where $x^{(s)}$ and $y^{(s)}$ are the coordinates of sensor $s$.
The range-bearing measurement model is
\[
\mathbf{z}_{s,k}= \begin{bmatrix}
\sqrt{( x_k \!-\! x^{(s)} )^2 + (y_k \!-\! y^{(s)})^2} \,\,\\[.5mm]
\tan^{-1}\!\Big( \frac{ x_k - x^{(s)} }{ y_k - y^{(s)} } \Big) 
\end{bmatrix} 
+ \begin{bmatrix} v_{s,k}^{(1)} \\[1.5mm] v_{s,k}^{(2)} \end{bmatrix} \rmv,
\vspace{-.7mm}
\]
where $v_{s,k}^{(1)}$ and $v_{s,k}^{(2)}$ are, individually, iid zero-mean Gaussian with standard deviation $\sigma_1 \!=\! 20\ist$m and 
$\sigma_2 \rmv=\rmv (\pi/90)\ist$rad, \nolinebreak 
respectively. \nolinebreak 
The field of view of the nonlinear sensors is a disc\linebreak 
of radius 3000m centered at the sensor position; this disc always covers the entire ROI. 
For both the linear and the \nolinebreak 
nonlinear \nolinebreak 
sensors, clutter is uniformly distributed over the sensor's field of view with an average number of ten clutter measurements per time step, 
or equivalently clutter intensity 
$\kappa_{s,k}(\mathbf{z}_k)$\linebreak 
$=\rmv 10/(2000^2) \rmv=\rmv 2.5 \cdot 10^{-6}$ for the linear sensors and 
$\kappa_{s,k}(\mathbf{z}_k)$\linebreak 
$=\rmv 10/(2\pi \rmv\cdot\rmv 3000) \approx 5.31 \cdot 10^{-4}$ 
for the nonlinear sensors. The clutter measurements of different sensors are independent. 


\subsubsection{Local PHD Filters} 
\label{sec:SimulationSet_filters}
We consider two scenarios. In the first scenario, all the local PHD filters use a particle-based implementation.
In the second scenario, only the local PHD filters at the nonlinear sensor nodes use a particle-based implementation, whereas the local PHD filters at the linear sensor nodes 
use a GM-based implementation \cite{Vo06,Li17PC}. 
The results for the second scenario demonstrate the applicability of our distributed PHD filter in heterogeneous networks combining
particle-based and GM-based local PHD filters. 


We compare the performance and computing time of the following particle-based PHD 
\vspace{.5mm}
filters:

\begin{itemize}

\item The proposed distributed PHD filter, which will be briefly referred to as \emph{AA-F-IS} or \emph{AA-C-IS} depending on whether flooding (F) or average consensus (C)
is used as the dissemination/fusion scheme.

\vspace{1mm}

\item  
A modified version of the GA fusion-based, particle-based, distributed PHD filter proposed in \cite{Uney13}, briefly referred to as \emph{GA-EMD}.
In 
\cite{Uney13}, two important steps are\linebreak 
(i) a conversion of the particle representation of the PHD\linebreak 
into a kernel-based 
representation, and (ii) the construction of the 
multitarget exponential mixture density (EMD). Regarding the first step, we replaced the clustering algorithm for kernel function learning proposed in \cite{Uney13}---which 
we observed in our simulations to be computationally intensive and potentially unstable---with our particles-GM conversion algorithm from Section \ref{sec:P2GM}.
Regarding the second step, we use
our IS method for GM-particles conversion (see Section \ref{sec:IS}) for updating the fused particles.
Finally, we do not employ the sophisticated strategy for online adjustment of the fusion weights proposed in \cite{Uney13} but use fixed Metropolis weights, 
which have been widely used for GA-based GM-PHD 
fusion \cite{Battistelli13}.

\vspace{1mm}

The resulting modification of the EMD fusion method of
\cite{Uney13}
is more computationally efficient, although---as shown later---it is still considerably less efficient than our proposed fusion method. 
Moreover, just as the filter of
\cite{Uney13}, it has
a significantly higher
communication cost\linebreak 
because it communicates both the particles 
and 
the kernel/GM parameters. For this reason, using flooding for dissemination/fusion is infeasible, and hence we only use the average consensus scheme.

\vspace{1mm}

\begin{figure*}
\vspace{-2mm}
\centering
\includegraphics[width=18 cm]{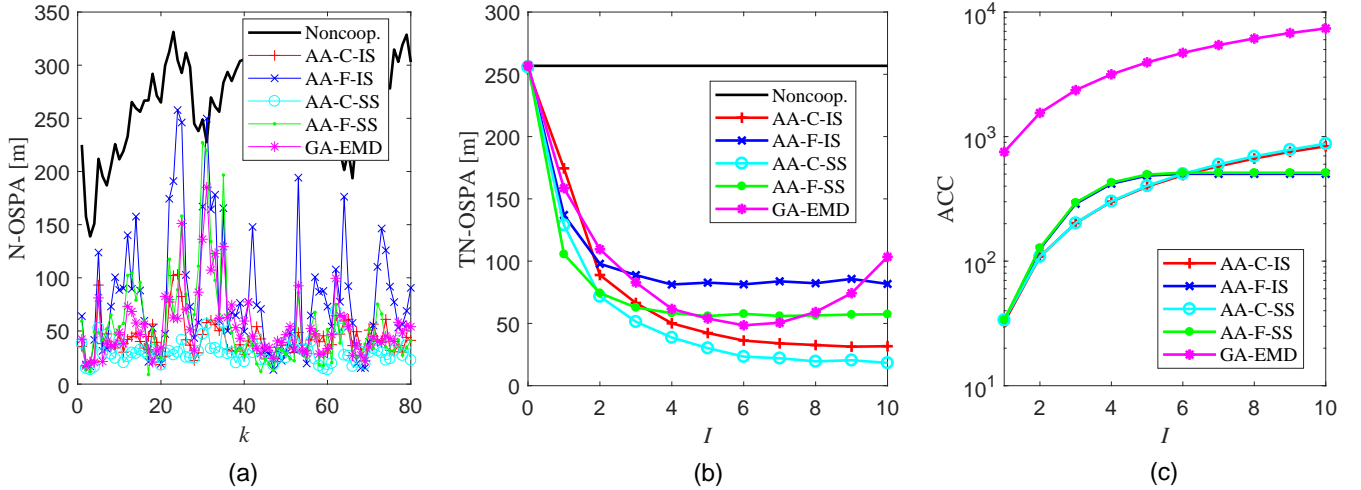}
\vspace{-1.5mm}
\caption{Results for the first scenario: 
(a) Network OSPA error 
versus time $k$ (here, the distributed filters use $I \!=\! 5$ dissemination/fusion iterations). 
(b) Time-averaged network OSPA error versus number of dissemination/fusion iterations $I$. 
(c) Average communication cost versus $I$.} 
\label{fig:scenario_1} 
\end{figure*}

\item 
A modified version of our proposed distributed PHD filter, in which the 
GM--particles conversion is done via
the standard sampling (SS) method---i.e., sampling directly from the fused GM-PHD 
$D_{s,k}^{\text{GM}[I]}(\mathbf{x})$---instead of our IS method from Section \ref{sec:IS}. We consider this filter 
to 
compare the IS method with the SS method. 
We 
refer to it
as \emph{AA-F-SS} or \emph{AA-C-SS} depending on the dissemination/fusion scheme employed. 



\vspace{1mm}

\item 
A noncooperative PHD filter in which each local PHD filter relies solely on its own local measurements and does not communicate with other local PHD filters.

\vspace{1mm}

\vspace{1mm}

\end{itemize}


The local PHD filters use systematic resampling \cite{Li15SPM}, and\linebreak 
they adjust the number of particles via resampling to be $200\cdot\rmv \hat{N}_{s,k}^{\text{local}}$ 
if $\hat{N}_{s,k}^{\text{local}}\!\geq\rmv 0.5$ and 100 otherwise, where $\hat{N}_{s,k}^{\text{local}}\rmv= \mathrm{round}\{ W_{s,k}\}$. 
(Here, we use $ W_{s,k}$ and not $W_{\! s,k}^{[I]}$ because in the resampling step, $W_{\! s,k}^{[I]}$ is not available yet.)
\vspace{-.5mm} 
The target state estimates $\hat{\mathbf{x}}_k^{(\nu)}$
are calculated
as 
described in Section \ref{sec:EstimateExtraction}.
The threshold defining 
$\Omega_{s,k}(\mathbf{z})$ (see Section \ref{sec:P2GM}) is $T_\Omega \!=\rmv 0.3$.
The consensus-based filters (AA-C-SS/IS and GA-EMD) perform GC merging in each consensus iteration 
(see Section \ref{sec:P2GM_merg}); GCs are merged if their Mahalanobis distance is smaller than 2 \cite{Salmond90}.

%
For each of the two scenarios,
we performed 100 simulation runs using the target trajectories shown in Fig.\ \ref{fig:scene1} and 
randomly generated measurement noise and initial particles. Each simulation run consists of 80 time steps. 

\vspace{-1mm}

\subsection{First Scenario---Particle-based Local PHD Filters}
\label{sec:particle_PHD_scene1}



In the first scenario, all the local PHD filters use a particle-based implementation. 

\subsubsection{Tracking Accuracy}
\label{sec:particle_PHD_scene1_ospa}
%
We quantify the target detection and position estimation performance of the 
filters by the mean optimal subpattern assignment (OSPA) error
\cite{Schuhmacher08} with cutoff parameter $c \rmv=\! 1000\ist$m and order 
$p \rmv=\rmv 2$. More specifically, we consider the average of the OSPA errors obtained by all the sensors,
referred to as \textit{network OSPA error} (briefely N-OSPA) and the average of the network OSPA errors over all the 80 time steps, referred to as 
\textit{time-averaged network OSPA error} (TN-OSPA). Fig.~\ref{fig:scenario_1}(a) shows the N-OSPA of the 
distributed PHD filters using $I \!=\! 5$ 
dissemination/fusion iterations, as well as of the noncooperative PHD filter, versus time $k$. Fig.~\ref{fig:scenario_1}(b) shows the TN-OSPA 
versus the number $I$ of dissemination/fusion iterations. One can see that the D-PHD filters have a significantly smaller OSPA error than the noncooperative PHD filter. 

According to Fig.~\ref{fig:scenario_1}(b), the reduction of the TN-OSPA for growing $I$ is quite fast initially.
For larger $I$, the TN-OSPA decreases more slowly (in the case of the consensus-based filters) or it stays roughly constant (in the case of the flooding-based filters),
or it even starts increasing again (in the case of GA-EMD). Regarding the flooding-based filters, we recall from Section \ref{sec:P2GM_flood} that the flooding dissemination 
of the GM parameters is already complete when $I$ equals the network diameter $R\rmv=\rmv 5$, and thus no further gains can be achieved for $I\rmv\ge\rmv 6$.
Furthermore, we conjecture that the increase of the TN-OSPA of GA-EMD
for $I \rmv\ge\rmv 7$ is due to the fact that a missed detection at any single sensor
can degrade the performance of GA fusion significantly, and the probability of such a missed detection 
increases when more sensors are involved. We note that a similar increase of the OSPA for additional GA dissemination/fusion iterations
was reported in \cite[Fig.~8]{Battistelli15} (in the intervals $k \!\in\! [25\ist\text{s}, 100\ist\text{s}]$ and $k \!\in\! [790\ist\text{s}, 800\ist\text{s}]$).
It is furthermore seen in Fig.~\ref{fig:scenario_1}(b) that the TN-OSPA of GA-EMD is larger than that of AA-C-IS/SS 
(except for $I \!=\! 1$, where according to Fig.~\ref{fig:scenario_1}(b) it is slightly smaller than that of AA-C-IS).

The OSPA performance of the SS-based filters (AA-F-SS\linebreak 
and AA-C-SS) is seen to be better than that of the corresponding IS-based filters 
(AA-F-IS and AA-C-IS, respectively). This is because
sampling directly from $D_{s,k}^{\text{GM}[I]}(\mathbf{x})$
represents $D_{s,k}^{\text{GM}[I]}(\mathbf{x})$ more accurately than the indirect sampling performed by our IS method.
(However, we recall
that the IS method
enables the far-reaching parallelization of filtering and fusion operations 
described in Section \ref{sec:summ-parallel}.)
Finally, the consensus-based filters (AA-C-SS and AA-C-IS) outperform the flooding-based filters (AA-F-SS and AA-F-IS, respectively); the only exception is $I \!=\! 1$, where
the consensus and flooding schemes differ merely by the choice of the fusion weights (uniform and Metropolis weights, respectively).
This 
superiority of the GM consensus scheme (for $I \rmv\ge\rmv 2$) is unexpected,
since flooding yields a faster dissemination of the GM parameters than consensus. A possible reason is
the GC merging 
performed by the GM consensus scheme in each fusion iteration.
In this context, an interesting observation is that GA-EMD---which is also consensus-based and performs GC merging---outperforms AA-F-IS for $3 \!\leq\! I \!\leq\! 9$. \nolinebreak 
In \nolinebreak 
additional simulations for various scenarios,
we observed that the performance of consensus-based PHD filter algorithms with GC merging, including 
AA-C-SS/IS and GA-EMD, is highly sensitive to the threshold of the Mahalanobis distance used for GC merging: we found that threshold 2 yields the best 
filter performance, whereas other thresholds 
can lead to a significantly poorer 
performance. 
   

\subsubsection{Communication Cost}
\label{sec:particle_PHD_scene1_comm}
We measure the \emph{average communication cost} (ACC) of the various filters by the 
number of real values broadcast 
by a sensor to its neighbors during all the dissemination/fusion iterations performed at one time step, 
averaged
over all the sensors, time steps, and simulation runs. Note that in addition to one real value for the cardinality estimate, only GC parameters are broadcast 
in AA-F/C-IS and AA-F/C-SS whereas in GA-EMD, both GC parameters and unweighted particles (i.e., the particles after the resampling step) 
are broadcast. Here, each unweighted particle amounts to four real values. 

Fig.~\ref{fig:scenario_1}(c) shows the ACC
versus 
$I$. The increase of the ACC\linebreak 
of GA-EMD and AA-C-IS/SS with $I$ is an expected result because 
the ACC was defined as the average total communication cost for all the $I$
dissemination/fusion iterations.
The ACC of AA-F-IS/SS increases 
up to $I \!=\! 5$ but stays constant afterwards. This is also expected because, as mentioned earlier,
the flooding dissemination is already complete when $I \!=\! R\rmv=\rmv 5$, and thus no additional information needs to be communicated
for $I\rmv\ge\rmv 6$.
The ACC of GA-EMD is seen to be larger by about one order of magnitude than that of the other filters; this is 
because GA-EMD communicates a large number of particles in addition to GC parameters. Furthermore, the ACC of the flooding-based filters 
is larger than that of the consensus-based filters for $I$ between 2 and 5, and smaller for $I \geq 7$. 
At this point, we recall from Section \ref{sec:comm_cost}
that the communication cost of the consensus-based filters strongly depends on the 
GC merging.
Using a larger threshold for the Mahalanobis distance (so that more GCs are merged) would result in 
a smaller communication cost 
but also in a poorer tracking accuracy.
Finally, AA-F-SS and AA-F-IS are seen to have almost the same ACC, and similarly for AA-C-SS and AA-C-IS. 
This is because the choice of the GM--particles conversion method---SS or IS---has only little effect on the communication cost.



\begin{table}[t!]
\caption{Results for the first scenario: Average computing time of one filtering step. The distributed filters use $I \!=\! 5$ dissemination/
fusion iterations.}
\vspace{-2.5mm}
\label{tab:1}
\begin{center}
\begin{tabular}{|c|c|}
\hline
\rule[-1.3mm]{0mm}{4.2mm}Filter & Average\hspace{.7mm}Computing{\,}Time{\,}[s]\\
\hline
\hline
\rule[-1.3mm]{0mm}{4.2mm}Noncooperative	&0.079 \\
\hline
\rule[-1.3mm]{0mm}{4.2mm}AA-F-SS 	&0.181 \\
\hline
\rule[-1.3mm]{0mm}{4.2mm}AA-C-SS 	&0.347 \\
\hline
\rule[-1.3mm]{0mm}{4.2mm}AA-C-IS 	&0.387 \\
\hline
\rule[-1.3mm]{0mm}{4.2mm}AA-F-IS 	&0.558 \\
\hline
\rule[-1.3mm]{0mm}{4.2mm}GA-EMD	& 1.837 \\
\hline
\end{tabular}
\end{center}
\vspace{-2mm}
\end{table}

\subsubsection{Computational Complexity}
\label{sec:particle_PHD_scene1_compl}
Finally, 
we quantify the computational complexity of the various filters by the \emph{average computing time} of each filtering step (corresponding to each time step $k \to k \rmv+\! 1$),
where the averaging is over all the local PHD filters, time steps, and simulation runs. 
The computing times were obtained using MATLAB implementations on an Intel Core M-5Y71 CPU. 
Table~\ref{tab:1} shows the average computing time 
for the 
distributed PHD filters using $I \!=\! 5$ 
dissemination/fusion iterations, as well as for the noncooperative PHD filter, versus time $k$. 
It is seen that GA-EMD is significantly more complex than 
the other distributed filters. Furthermore, AA-F-IS and AA-C-IS are more complex than AA-F-SS and AA-C-SS; this
is because the IS method is 
more complex
than the SS method. 
AA-F-IS is more complex than AA-C-IS, due to the larger number of GCs that are processed.
Indeed, in AA-C-IS, the number of GCs is reduced by GC merging,
and the complexity of the GM merging operations is considerably smaller than the added complexity 
of AA-F-IS caused by the additional GCs. On the other hand, AA-F-SS is less complex than AA-C-SS. Here, the reason is that the SS method employed by
AA-F-SS and AA-C-SS 
has a low complexity,
and thus the complexity of the merging operations performed by AA-C-SS is larger than the added complexity of AA-F-SS caused by the additional GCs. 

\vspace{-1mm}
 
\subsection{Second Scenario---Particle-based and GM-based Local PHD Filters} 
\label{sec:particle_PHD_scene_hybrid}

\begin{figure*}
\vspace{-2mm}
\centering
\includegraphics[width=18 cm]{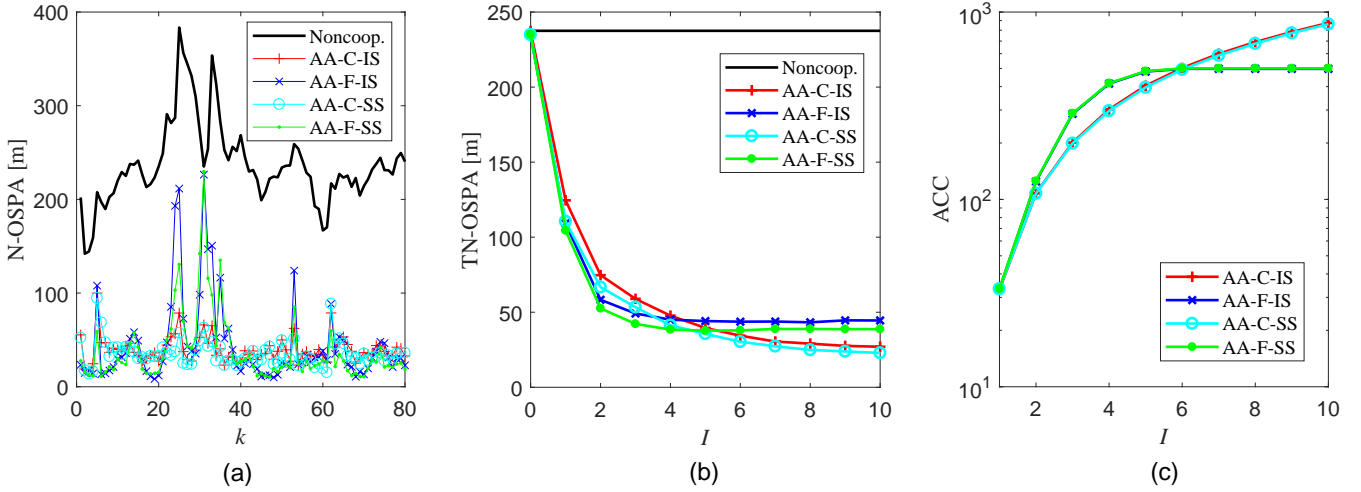}
\vspace{-1.5mm}
\caption{Results for the second scenario: 
(a) Network OSPA error 
versus time $k$ (here, the distributed filters use $I \!=\! 5$ dissemination/fusion iterations). 
(b) Time-averaged network OSPA error versus number $I$ of dissemination/fusion iterations. 
(c) Average communication cost versus $I$.} 
\label{fig:scenario_2} 
\end{figure*}

Next, we study 
a heterogeneous network where the eight nonlinear sensor nodes use a 
particle-based local PHD filter and the eight linear sensor nodes use a GM-based local PHD filter \cite{Vo06,Li17PC} (briefly referred to as GM-PHD filter).
The sensor network topology and the target trajectories are as before (see Fig.~\ref{fig:scene1}).
The GM-PHD filters use at most 100 GCs. For mixture reduction, following \cite{Vo06},
they remove GCs with a weight smaller than $10^{-4}$ and merge GCs \nolinebreak 
with \nolinebreak 
a \nolinebreak 
Maha\-la\-nobis distance smaller than 4.
(We note that here, the Mahalanobis distance threshold 4 performed better than the threshold 2 that\linebreak 
we used in the consensus-based particle-PHD filters in Section 
\ref{sec:particle_PHD_scene1}.)
Furthermore, for fusing their local GM with the GMs of the other sensors, the GM-PHD filters perform a straightforward union of the GM parameter sets 
and subsequently adjust the weights using the cardinality averaging method discussed in Section \ref{sec:CC-AA}.
The combination---within the sensor network---of the GM-PHD filters with the particle-based AA-F/C-SS/IS filters will be briefly referred to as ``AA-F/C-SS/IS.'' 
We no longer consider GA-EMD as it cannot be\linebreak 
combined with a GM-PHD filter 
in a straightforward fashion (i.e., without conversions between particle and
GM representations).

The simulation results for this scenario, shown in Fig.~\ref{fig:scenario_2} and Table~\ref{tab:2}, are generally similar to those for the first scenario
(see Fig.~\ref{fig:scenario_1} and Table~\ref{tab:1}). A difference is that now AA-F-IS and AA-F-SS have a smaller TN-OSPA than, respectively, 
AA-C-IS and AA-C-SS for $I \!\leq\rmv 4$, instead of only for $I \!=\! 1$ (as was the case in the first scenario). 
This is because now 
half of the local filters are 
GM-PHD filters, 
for which flooding-based fusion performs better than consensus-based fusion \cite{Li17PC}.

\begin{table}[t!]
\caption{Results for the second scenario: Average computing time of one filtering step. The distributed filters use $I \!=\! 5$ 
dissemination/fusion iterations.}
\vspace{-2.5mm}
\label{tab:2}
\begin{center}
\begin{tabular}{|c|c|}
\hline
\rule[-1.3mm]{0mm}{4.2mm}Filter & Average\hspace{.7mm}Computing{\,}Time{\,}[s]\\
\hline
\hline
\rule[-1.3mm]{0mm}{4.2mm}Noncooperative	&0.096 \\
\hline
\rule[-1.3mm]{0mm}{4.2mm}AA-F-SS 	&0.233 \\
\hline
\rule[-1.3mm]{0mm}{4.2mm}AA-C-SS 		&0.353 \\
\hline
\rule[-1.3mm]{0mm}{4.2mm}AA-C-IS 	&0.381 \\
\hline
\rule[-1.3mm]{0mm}{4.2mm}AA-F-IS 	&0.422 \\
\hline
\end{tabular}
\end{center}
\vspace{-3mm}
\end{table}

\section{Conclusion}  
\label{sec:conclusion}

We proposed a distributed PHD (D-PHD) filter where the local 
filters use a particle-based implementation 
to support 
nonlinear/non-Gaussian 
system models, but the 
fusion of the local PHDs is based on a Gaussian mixture (GM) representation 
to reduce
communication and enable an easy combination with GM-based local 
filters. Our D-PHD filter differs from most
existing
filters in that it
seeks to compute the arithmetic average (AA) of the local PHDs, 
rather than 
the geometric average (GA).
Two noteworthy components of our\linebreak 
D-PHD filter algorithm are (i) a ``significance-based'' method for converting 
particle representations into GM representations, which reduces communication 
and complexity,
and (ii) an importance sampling method for converting the fused GMs 
into particle representations, which enables a parallelization of 
filtering 
and fusion operations. This parallelization is especially advantageous when the sensing rate is high and/or the 
duration of 
one dissemination/fusion iteration is large.

An experimental comparison of our 
filter with a state-of-the-art 
filter using GA fusion showed that, in the considered scenarios, 
consensus-based AA fusion outperforms consensus-based GA fusion in terms of estimation accuracy, 
complexity, and communication cost. 
Our simulations
also showed that consensus-based AA fusion 
can 
outperform flooding-based AA fusion
in terms of both estimation accuracy and communication cost. We 
expect that this 
advantage of AA fusion can to be further increased by using more sophisticated mixture reduction schemes such as \cite{Crouse11,Ardeshiri15}.


\appendices
\renewcommand*\thesubsectiondis{\thesection.\arabic{subsection}}
\renewcommand*\thesubsection{\thesection.\arabic{subsection}}

\vspace{-1mm}

\section{Proof of Eq.\ \eqref{eq:ResamplingProposal}} 
\label{appendix:A}

To show that the choice of $P_{j'}$ in \eqref{eq:ResamplingProposal} yields \eqref{eq:E_resampling}, we note that $N_{s,k}^{(j')}$ can be written as
$N_{s,k}^{(j')} \!=\rmv \sum_{j=1}^{\tilde{J}_{s,k}} I\big[\tilde{\mathbf{x}}_{s,k}^{(j)} \!=\rmv \mathbf{x}_{s,k}^{(j')}\big]$,
where $I\big[\tilde{\mathbf{x}}_{s,k}^{(j)} \!=\rmv \mathbf{x}_{s,k}^{(j')}\big]$ equals $1$ if 
$\tilde{\mathbf{x}}_{s,k}^{(j)} \!=\rmv \mathbf{x}_{s,k}^{(j')}$ and $0$ otherwise.
Thus, 
\vspace{-2.5mm}
\be
\mathrm{E} \big[N_{s,k}^{(j')} \big| \xi_{s,k} \big] = \sum_{j=1}^{\tilde{J}_{s,k}} \mathrm{E} \big[ I\big[\tilde{\mathbf{x}}_{s,k}^{(j)} \!=\rmv \mathbf{x}_{s,k}^{(j')}\big] \big| \xi_{s,k} \big] \ist.
\label{eq:sum_E}
\vspace{-2mm}
\ee
Now 
\vspace{-.5mm}
\begin{align*}
&\mathrm{E} \big[ I\big[\tilde{\mathbf{x}}_{s,k}^{(j)} \!=\rmv \mathbf{x}_{s,k}^{(j')}\big] \big| \xi_{s,k} \big] \\[0mm]
&\;\; = 1 \cdot \mathrm{Pr}\big[\tilde{\mathbf{x}}_{s,k}^{(j)} \!=\rmv \mathbf{x}_{s,k}^{(j')}  \big| \xi_{s,k} \big] + 0 \cdot \mathrm{Pr}\big[\tilde{\mathbf{x}}_{s,k}^{(j)} \!\not=\rmv \mathbf{x}_{s,k}^{(j')}  \big| \xi_{s,k} \big] \\[0mm]
&\;\; = P_{j'} \ist,
\end{align*}
whence \eqref{eq:sum_E} 
\vspace{-2mm}
becomes
\begin{align*}
\mathrm{E} \big[N_{s,k}^{(j')} \big| \xi_{s,k} \big] &= \sum_{j=1}^{\tilde{J}_{s,k}} P_{j'} \nn \\[.4mm]
&= \tilde{J}_{s,k} \ist  P_{j'}\nn \\[.4mm]
&=\mathrm{round} \{ N_{\textrm{p}} \ist W_{s,k} \} \ist  P_{j'} \nn \\[.4mm]
&\approx N_{\textrm{p}} \ist W_{s,k} \ist P_{j'} \nn \\[0mm]
&= N_{\textrm{p}}\ist w_{s,k}^{(j')} ,
\end{align*}
where \eqref{eq:Jsk_round} and \eqref{eq:ResamplingProposal} have been used. Hence, to within a rounding error (caused by replacing $\mathrm{round} \{ N_{\textrm{p}} \ist W_{s,k} \}$ 
with $N_{\textrm{p}} \ist W_{s,k}$), $\mathrm{E} \big[N_{s,k}^{(j')} \big| \xi_{s,k} \big]$ equals $N_{\textrm{p}}\ist w_{s,k}^{(j')}$, as postulated in \eqref{eq:E_resampling}.

\bibliographystyle{IEEEtran}

\bibliography{ISsmcGM}

\end{document}